\documentclass[aps,prd,groupedaddress,preprint,eqsecnum,%
nofootinbib,showpacs
]{revtex4}
\usepackage{graphicx,epsf,amssymb,amsbsy,amsfonts,amssymb,amsmath}
\usepackage[usenames, dvipsnames]{color}
\usepackage{calc}
\usepackage{url}
\usepackage{array}

\newif\ifdraft
\drafttrue
\newif\ifpreprint
\preprinttrue

\def\sect#1{section~{\ref{#1}}}
\def\fig#1{fig.~{\ref{#1}}}

\def\eqn#1{eq.~(\ref{#1})}
\def\Eqn#1{Equation~(\ref{#1})}
\def\eqns#1#2{eqs.~(\ref{#1}) and~(\ref{#2})}

\def\be{\begin{equation}}
\def\ee{\end{equation}}
\def\bea{\begin{eqnarray}}
\def\eea{\end{eqnarray}}
\def\ba{\begin{eqnarray}}
\def\ea{\end{eqnarray}}
\def\spa#1.#2{\left\langle#1\,#2\right\rangle}
\def\spb#1.#2{\left[#1\,#2\right]}
\def\spash#1.#2{\spa{\smash{#1}}.{\smash{#2}}}
\def\spbsh#1.#2{\spb{\smash{#1}}.{\smash{#2}}}
\def\sand#1.#2.#3{%
\left\langle\smash{#1}{\vphantom1}^{-}\right|{#2}%
\left|\smash{#3}{\vphantom1}^{-}\right\rangle}
\def\sandpp#1.#2.#3{%
\left\langle\smash{#1}{\vphantom1}^{+}\right|{#2}%
\left|\smash{#3}{\vphantom1}^{+}\right\rangle}
\def\sandpm#1.#2.#3{%
\left\langle\smash{#1}{\vphantom1}^{+}\right|{#2}%
\left|\smash{#3}{\vphantom1}^{-}\right\rangle}
\def\sandmp#1.#2.#3{%
\left\langle\smash{#1}{\vphantom1}^{-}\right|{#2}%
\left|\smash{#3}{\vphantom1}^{+}\right\rangle}
\def\sandmm#1.#2.#3{%
\left\langle\smash{#1}{\vphantom1}^{-}\right|{#2}%
\left|\smash{#3}{\vphantom1}^{-}\right\rangle}
\def\spab#1.#2.#3{\sandmm#1.#2.#3}
\def\spaa#1.#2.#3.#4{\sandmp#1.{#2#3}.#4}
\def\spbb#1.#2.#3.#4{\sandpm#1.{#2#3}.#4}

\def\Tr{\, {\rm Tr}}

\def\fs#1{{      
        \setbox\charbox=\hbox{$#1$}
        \setbox\slabox=\hbox{$/$}
        \dimen\charbox=\ht\slabox
        \advance\dimen\charbox by -\dp\slabox
        \advance\dimen\charbox by -\ht\charbox
        \advance\dimen\charbox by \dp\charbox
        \divide\dimen\charbox by 2
        \raise-\dimen\charbox\hbox to \wd\charbox{\hss/\hss}
        \llap{$#1$} }}
\def\ksl{\fs{k}}

\def\tf{\tilde{f}}
\def\tA{\tilde{A}}
\def\subtractfour#1{\ifthenelse{#1=5}{1}{\ifthenelse{#1=6}{2}
{\ifthenelse{#1=7}{3}{\ifthenelse{#1=8}{4}{\ifthenelse{#1=9}{5}
{\ifthenelse{#1=10}{6}{\ifthenelse{#1=11}{7}{\ifthenelse{#1=12}{8}
{\ifthenelse{#1=13}{9}{\ifthenelse{#1=14}{10}{}}}}}}}}}}}
\def\a{\alpha}
\def\b{\beta}

\def\k{\kappa}
\def\m{\mu}
\def\n{\nu}
\def\r{\rho}

\def\s{\sigma}

\def\p{\phi}

\def\la{\lambda}

\newcommand{\ad}{{\dot{\alpha}}}

\newcommand{\CA}{\mathcal{A}}       

\newcommand{\CP}{\mathcal{P}}       
\newcommand{\CN}{\mathcal{N}}      
\newcommand{\CO}{\mathcal{O}}      
\def\MHV{{\rm MHV}}
\def\MHVbar{$\overline{\hbox{MHV}}$}

\def\YM{{\rm YM}}

\def\hf{{\textstyle{\frac{1}{2}}}}
\def\ihf{{\textstyle{\frac{i}{2}}}}
\def\sst{\scriptscriptstyle}

\newbox\charbox
\newbox\slabox

\newcommand{\nn}{\nonumber}
\newcommand{\nnl}{\nonumber\\}

\def\_{\;\;}
\def\^{\wedge}

\def\diff{\mbox{d}}

\def\pd{\mbox{$\partial$}}
\def\sfrac#1#2{{\textstyle\frac{#1}{#2}}}
\def\>{\rangle}
\def\<{\langle}
\def\+{\dagger}
\def\={\ =\ }
\def\and{\qquad\textrm{and}\qquad}

\begin{document}
\hfuzz 20pt

\noindent
SU-ITP-11/51$\null\hskip 10 truecm\null$ SLAC--PUB--14784
\vskip1cm

\title{Color-kinematics duality and double-copy construction\\
for amplitudes from higher-dimension operators}

\author{Johannes Broedel${}^{1}$ and Lance J. Dixon${}^{2}$}

\affiliation{
${}^1$Stanford Institute for Theoretical Physics and
Department of Physics, Stanford University,
Stanford, CA 94305, USA\\
${}^2$SLAC National Accelerator Laboratory, 
Stanford University, Stanford, CA 94309, USA}

\begin{abstract}
We investigate color-kinematics duality for gauge-theory amplitudes
produced by the pure nonabelian Yang-Mills action deformed by
higher-dimension operators.  For the operator denoted by $F^3$,
the product of three field strengths, the existence of color-kinematic
dual representations follows from string-theory monodromy relations.
We provide explicit dual representations, and show how
the double-copy construction of gravity amplitudes based on them 
is consistent with the Kawai-Lewellen-Tye relations.  It leads to
the amplitudes produced by Einstein gravity coupled to a dilaton field
$\phi$, and deformed by operators of the form $\phi R^2$ and $R^3$.
For operators with higher dimensions than $F^3$, such as $F^4$-type
operators appearing at the next order in the low-energy expansion of 
bosonic and superstring theory, the situation is more complex.
The color structure of some of the $F^4$ operators is incompatible with a 
simple color-kinematics duality based on structure constants $f^{abc}$, but
even the color-compatible $F^4$ operators do not admit the duality.
In contrast, the next term in the $\a'$ expansion of the superstring
effective action --- a particular linear combination of $D^2F^4$ and
$F^5$-type operators --- does admit the duality, at least for amplitudes
with up to six external gluons.
\end{abstract}

\pacs{04.60.-m, 04.60.Cf, 11.15.Bt, 11.55.Bq \hspace{1cm}}

\maketitle
\thispagestyle{empty}
\tableofcontents

\pagestyle{plain}
\setcounter{page}{1}


\section{Introduction}\label{sec:Introduction}

In the 1980s, Kawai, Lewellen and Tye (KLT)~\cite{KLT} found a relation
between tree-level open and closed string tree-level amplitudes,
which implies --- after taking the field-theory limit --- that the 
color-ordered tree-level scattering amplitudes for gluons in Yang-Mills (YM)
theories contain all the information necessary to construct the tree-level 
graviton scattering amplitudes in suitable gravity theories.  More recently,
Bern, Carrasco and Johansson (BCJ)~\cite{BCJ} discovered that it is possible 
to directly construct gravity amplitudes from gauge-theory amplitudes,
if the latter can be organized so that their numerator factors respect
a certain color-kinematics duality.  This method for obtaining gravity
amplitudes is known as the double-copy construction.
Color-kinematics duality is associated with linear BCJ relations between
color-ordered subamplitudes.  The BCJ amplitude relations have
been proven using string-theory
monodromy relations~\cite{MonodromyNBI1,MonodromyStieberger,MonodromyNBI2}
and directly in field theory~\cite{BoFengBCJProof}
using the BCFW recursion relations~\cite{BCFW}.
Given a color-kinematic dual
representation for gauge theory, the double-copy construction has been
proven to hold for all tree-level gravity amplitudes by Bern, Dennen,
Huang, and Kiermaier~\cite{LagrangeBCJ}.  Furthermore, in
contrast to the field-theory limit of the KLT relations, the double-copy
construction has been conjectured to hold for integrands of loop-level
amplitudes as well.  This conjecture has been verified, at different levels,
in numerous examples~\cite{BCJ,LoopBCJ,loop5pt,JJHenrikReview,%
BBJ,BoucherDixon,SchnitzerBCJ,FourLoopBCJ}.

Color-kinematic dual representations exist for a variety of
gauge-theory amplitudes.
They have been shown to exist explicitly for tree-level amplitudes with
up to eight external legs~\cite{BCJ}.  Expressions are also
available for several loop-level integrands in maximally supersymmetric
Yang-Mills (sYM) theory~\cite{BCJ,LoopBCJ,loop5pt,FourLoopBCJ},
supporting the conjecture that all amplitudes of the theory can be cast
into this form.

Gauge theories with fewer than the maximal number of supersymmetries
are also expected to have dual representations, at both tree and
loop level.  At tree level, this assertion can often be demonstrated
using the fact that the tree amplitudes coincide with the maximally
supersymmetric case.  At loop level,
little has been shown explicitly to date (see, however,
ref.~\cite{BCJ} for a two-loop QCD example), primarily because the
loop integrands for amplitudes in gauge theories with fewer
supersymmetries are considerably more complicated than in the
maximally supersymmetric case.

All of these discussions of color-kinematics duality are for 
amplitudes whose color decomposition can be expressed
entirely in terms of the three-index antisymmetric Lie
algebra structure constants $f^{abc}$ for some nonabelian gauge group.
Recently it has been pointed out that the notion of
color-kinematics duality can be generalized to three-algebras,
with a four-index antisymmetric structure constant $f^{abcd}$
appearing in the trilinear product of generators,
$[T^a,T^b,T^c] = f^{abc}_{~~~d} T^d$~\cite{Bargheer2012gv}.
This algebra appears in the Bagger-Lambert-Gustavsson
theory~\cite{BLG}, whose amplitudes exhibit this new color-kinematics
duality.  Ref.~\cite{Bargheer2012gv} also provides evidence for
an associated double-copy construction of the amplitudes of
maximal three-dimensional supergravity.

In four dimensions, gauge theories are renormalizable.
However, they live in a larger class
of nonrenormalizable theories, corresponding to the addition to the
Lagrangian of a large variety of higher-dimension, gauge-invariant
local operators built
out of the gauge fields (and matter fields, if present).
Some of these operators appear in the low-energy effective action
of either bosonic or supersymmetric open string theory.
In order to explore just how general the phenomenon of color-kinematics
duality is, it is quite natural to ask whether the scattering amplitudes
in these modified theories also have color-kinematic dual representations.

For pure gauge theory, the simplest candidate operator for modifying the YM
action is the well-known operator
$F^3 \equiv \Tr (F_\m^{~\n} F_\n^{~\r} F_\r^{~\m})$,
where $F_{\m\n}\equiv F_{\m\n}^a T^a$ is the gluon field strength.
This operator will be our prime example of a higher-dimension operator,
but it is also special in several ways.  It has
dimension six, the lowest possible dimension for a modification of pure gauge
theory.  In the low-energy limit, or $\alpha'$ expansion, of the effective
action for the open bosonic string, the operator $F^3$ arises as the first
correction to the usual YM Lagrangian $\Tr(F_{\m\n}F^{\m\n}$).
(See, for example, section 6.5 in ref.~\cite{Polchinski1}.)

The group theory structure implied by the $F^3$ modification mimics
that of the YM action, in the sense
that the three-gluon amplitudes in both cases are proportional to
the structure constant $f^{abc}$. In addition, the diagrams contributing to a
particular amplitude are --- up to changing the type of one vertex ---
topologically identical to the cubic graphs from pure YM-theory. Those two
features suggest that color-kinematics duality might extend easily from 
amplitudes in 
YM theory to amplitudes produced by $F^3$.  We use the string-theory
monodromy relations~\cite{MonodromyNBI1,MonodromyStieberger,MonodromyNBI2}
to derive the BCJ relations between color-ordered subamplitudes for $F^3$
amplitudes.  We also compute a variety of $F^3$ amplitudes with different
helicity structures and multiplicities, and verify the existence of
color-kinematic dual representations.

In QCD, the operator $F^3$ represents the only gauge-invariant modification
of the three-gluon coupling which survives for on-shell three-point
kinematics (with complex momenta)~\cite{BenincasaConde}.
New physics, such as gluon
compositeness or heavy colored states, could in principle produce this
operator.  There have been several phenomenological studies of its effects
on high-energy collisions~\cite{Simmons,Zeppenfeld,DixonShadmi}.
In pure-jet events at hadron colliders, the operator is difficult
to observe, because at tree level it produces helicity amplitudes that
are orthogonal to those of QCD, and so the interference correction vanishes,
until one reaches amplitudes with five external partons, corresponding
to three-jet events~\cite{DixonShadmi}.  At the three-point
level, QCD produces the (all-outgoing) helicity configurations $({-}{-}{+})$
and $({+}{+}{-})$, while the $F^3$ operator produces only $({-}{-}{-})$ and
$({+}{+}{+})$.  At the four-point level, tree-level
QCD produces only $({-}{-}{+}{+})$ 
(plus permutations), while $F^3$ produces $({-}{-}{-}{-})$, $({-}{-}{-}{+})$,
$({+}{+}{+}{-})$ and $({+}{+}{+}{+})$.  In contrast, the four-point
amplitudes involving a massive quark, such as $gg \rightarrow t\bar{t}$,
do allow for a leading-order interference between the QCD amplitudes
and those produced by the $F^3$ modification~\cite{Simmons}.

Several amplitudes for the scattering of massless partons using an
$F^3$-modified action have been computed
previously~\cite{DixonShadmi,DixonGloverKhoze,CEK}.  (Related amplitudes for a
Higgs boson $H$ coupled to gluons via a heavy top-quark loop and the operator
$HF^3$ have also been constructed recently~\cite{Neill}.) In
ref.~\cite{DixonGloverKhoze} it was proposed to use a modification of the rules
for pure gauge theory due to Cachazo, Svr\v{c}ek and Witten (CSW)~\cite{CSW},
which are called CSW rules and employ maximally-helicity-violating (MHV)
vertices.  In ref.~\cite{BoelsF3} these rules were proven, using canonical
transformations, and even allowing for additional coupling of $F^3$ to
a scalar (Higgs) field.   Here we compute explicitly higher-multiplicity
$F^3$ amplitudes requiring a number of MHV vertices, so that we can check
their compatibility with color-kinematics duality.

Another purpose of this article is to investigate whether the double-copy
construction for gravity can be extended to the case of gravity perturbed by
the operators up to $R^3$ (a suitable contraction of three Riemann tensors),
utilizing the amplitudes produced by $F^3$.  We find evidence that it can,
provided that intermediate exchanges of the dilaton/axion field of string
theory are properly taken into account. The connection between gravity
amplitudes from $R^3$ and gauge-theory amplitudes from $F^3$ via the KLT
relations at the three- and four-point level was studied
previously~\cite{BjB2003a,BjB2003b}.  It is simple to see that the
three-point amplitudes are related by `squaring'.  Beyond the three-point
level, however, one cannot avoid the introduction of the dilaton on the
closed-string side in order to reproduce the corrections to the string
amplitude.  
The $F^3$-modified amplitudes are order-$\a'$ corrections to gluon
scattering in open bosonic string theory.  When one applies a double-copy
construction based on a pair of $F^3$-modified amplitudes, that corresponds to
an order-$\a^{\prime 2}$ correction to graviton amplitudes in
closed string theory.
At order $\a^{\prime 2}$, the closed string effective action contains an
operator of the form $R^3$.  However, at order $\a'$ it also contains an
operator of the form $e^{-2\phi} R^2$, which can emit a dilaton $\phi$.  Two
insertions of this operator will contribute to amplitudes where the external
states are all gravitons --- an intermediate dilaton is created and destroyed
by the two operator insertions.  For low-multiplicity amplitudes, we will
demonstrate explicitly how the various contributions combine in the double-copy
construction.

At higher multiplicity, the general arguments for the double-copy
construction in Einstein gravity~\cite{LagrangeBCJ} rely on a 
color-kinematic dual representation of the gauge-theory amplitude.
This representation is 
satisfied by all the $F^3$ amplitudes we have constructed.  However, the
double-copy arguments~\cite{LagrangeBCJ} also rely on deformations of
both gravity and gauge-theory amplitudes by complex-momentum
shifts~\cite{BCFW}.  Generic amplitudes perturbed by higher-dimension operators
typically do not converge at large values of the shift.
Therefore we do not have a general argument for a double-copy construction
of the amplitudes produced by $R^3$ (plus $e^{-2\phi} R^2$).
On the other hand, we will provide evidence, for all helicities up 
to six external legs, in favor of the $R^3$ double-copy construction.  
For the case of identical-graviton helicities, we can use an all-line
complex-momentum shift~\cite{CEK} to determine the $R^3$ amplitudes. 
We will also perform indirect tests of the validity of the
double-copy construction, or alternatively a KLT construction of $R^3$
amplitudes, by testing their Bose symmetry under the exchange of 
identical-helicity gravitons, and their universal behavior in
collinear and soft limits.

Returning to the gauge-theory Lagrangian, we consider the addition of yet
higher-dimension pure Yang-Mills operators (dimension eight or more).
At dimension eight, there are four independent operators involving
four gluon field strengths ($F^4$-type operators).
Two linear combinations of these four operators appear in the low-energy
expansion of the open bosonic string and open superstring, respectively.
The amplitudes generated by these two linear combinations of operators
are readily available from string theory.  They allow for the testing of 
possible color-kinematic dual representations. 
We find that the bosonic and superstring amplitudes individually
correspond to operators whose color structure is incompatible with
the usual color-kinematics duality, simply because they cannot be
expressed in terms of structure constants $f^{abc}$ alone.
(In this paper we do not consider extensions of color-kinematics duality
to four-index antisymmetric structure constants~\cite{Bargheer2012gv}.)
The difference between the bosonic string and superstring amplitudes
actually has a color-compatible representation, and even obeys
a BCJ-like monodromy relation.  However, we find that the 
corresponding operator does not admit a color-kinematic
dual representation, for reasons we shall explain.

On the other hand, at the next order in $\a'$, we find that the operator
appearing at $\CO(\a'^3)$ in the superstring effective action is
color-compatible, and its amplitudes do satisfy color-kinematics duality
at the level of four, five and six external gluons.
Beyond this order in $\a'$, we tested the superstring four-point amplitudes
and did not find any dual representations.  This property can be
explained simply by the fact that the color structures of the amplitudes at
orders higher than $\CO(\a'^3)$ cannot be expressed in terms of structure
constants $f^{abc}$ alone.

The article is organized as follows.  In~\sect{sec:Background},
we review color-kinematics duality and BCJ relations, as well
as the KLT relations and the double-copy construction. Section~\ref{sec:F3} is
devoted to the $F^3$-modified theory.  It explains how the amplitudes in this
theory can be computed using CSW or MHV rules, and provides evidence 
that they obey the duality.
In~\sect{sec:R3}, we discuss the double-copy construction for
amplitudes produced by operators up to $R^3$ in gravity. 
In~\sect{sec:Beyond} we discuss the insertion of operators in 
gauge theory with dimension greater than six, paying particular attention
to their color structure. At dimension eight, we find no operators that 
are consistent with color-kinematics duality; but at dimension ten,
the $\CO(\a'^3)$ term in the superstring effective action appears
to be consistent.
Finally, in~\sect{sec:Conclusions} we summarize our
findings and point out avenues for future research. 

\section{Background}\label{sec:Background}

\subsection{Color-kinematics duality}
\label{ssec:BCJrep}

A fully color-dressed $m$-gluon tree amplitude in YM theory can be
decomposed as 
\be\label{colordecomposition}
\CA_m\ \equiv\
\CA_m(a_1,\ldots,a_m)\ =\ \sum_{\s\in S_m/Z_m}
 \Tr(T^{a_{\s(1)}}\ldots T^{a_{\s(m)}})
         A_m(\s(1),\ldots,\s(m)) \,,
\ee
where the $a_i$ are adjoint indices, the $A_m$ are color-ordered
subamplitudes, and the sum is over non-cyclic permutations
$S_m/Z_m = S_{m-1}$. We have set the gauge coupling $g$ to 1 for 
convenience.  The trace runs over the color generators $T^a$,
which are assumed to be in the fundamental representation of the gauge group.
The generators are normalized to $\Tr(T^aT^b)=\delta^{ab}$, and the
structure constants are defined by
\be
\tf^{abc}\ \equiv\ i\sqrt{2} f^{abc}\ =\ \Tr( [T^a, T^b] T^c ) \,.
\label{structureConstants}
\ee
Starting from the decomposition in \eqn{colordecomposition}, it takes only a
couple of steps to demonstrate the idea behind color-kinematic dual
representations: the color and the kinematic part of the amplitude are
brought into a form in which they behave identically under certain symmetry
operations. As an example, we will perform the necessary steps on the
four-point tree amplitude, which has the color decomposition, 
\bea
\CA_4(a_1,a_2,a_3,a_4) &=& \sum_{\sigma\in S_3}
 \Tr(\s(1),\ldots,\s(4))\,\,A_4(\s(1),\ldots,\s(4))\nnl
&=& \overleftrightarrow\Tr(1,2,3,4)\,A_4(1,2,3,4)
+ \overleftrightarrow\Tr(1,3,2,4)\,A_4(1,3,2,4)\nnl
&&\hskip0.0cm\null
+ \overleftrightarrow\Tr(1,2,4,3)\,A_4(1,2,4,3)\,.
\label{explicitPerms}
\eea
Here we have shortened the expression by using the reflection
symmetry of the subamplitudes, and by writing
$\Tr(T^{a_{i_1}} T^{a_{i_2}} T^{a_{i_3}} T^{a_{i_4}}) = \Tr(i_1,i_2,i_3,i_4)$
and $\Tr(i_1,i_2,i_3,i_4)+\Tr(i_4,i_3,i_2,i_1) =
\overleftrightarrow\Tr(i_1,i_2,i_3,i_4)$.

\begin{figure}
  \includegraphics[width=11cm]{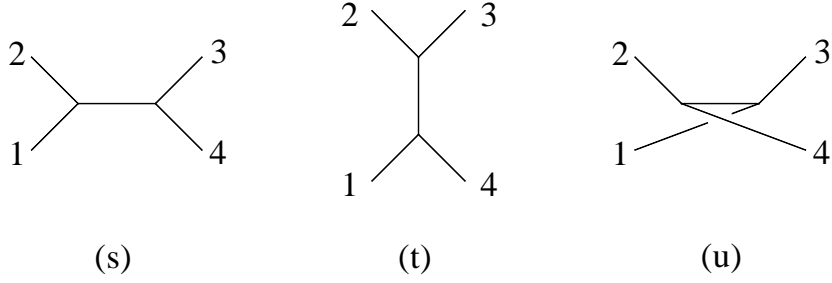}
  \caption{The three cubic graphs at the four-point level.}
  \label{fig:stugraphs}
\end{figure}

Color-kinematic
dual representations rely on the organization of amplitudes in terms
of cubic graphs.  For four external gluons, the three distinct cubic graphs
shown in \fig{fig:stugraphs} contribute to the scattering amplitude.
Each graph can be written as its color factor (here a product of two
structure constants) multiplied by a kinematic factor of the form
$\sfrac{\text{numerator}}{\text{propagator}}$,
which here takes the form, $\frac{n_s}{s}$, $\frac{n_t}{t}$
and $\frac{n_u}{u}$.
The Mandelstam variables appropriate for the four-point amplitude with
external momenta $k_1,k_2,k_3,k_4$ are $s=(k_1+k_2)^2$, $t=(k_2+k_3)^2$ and
$u=(k_1+k_3)^2$.  The numerator factors $n_i$ are labeled by the channel,
$s$, $t$ or $u$.

In a cubic-graph representation, the diagram produced by the four-gluon contact
interaction, which is proportional to the product of two structure constants,
has already been absorbed by expressing its kinematic part using inverse
propagators in the numerator, to compensate for the propagator denominators. 
(A similar construction is possible for other four-point contact interactions,
provided that they have the correct type of color structure.)

Each color-ordered amplitude can now be expressed in terms of two
cubic graphs.  For example, the first two graphs
in \fig{fig:stugraphs} contribute to $A_4(1,2,3,4)$,
\be
A_4(1,2,3,4)\ =\ \frac{n_s}{s}+\frac{n_t}{t}\,.
\label{4ptNumst}
\ee
Employing in addition the antisymmetry of the cubic vertices, which
implies, for instance,
\be
n_s(2,1,3,4) = - n_s(1,2,3,4) \,,
\label{CubicAntisymmetry}
\ee
we can rewrite \eqn{explicitPerms} as
\bea
\CA_4(a_1,a_2,a_3,a_4)
&=&
\bigl[ \overleftrightarrow\Tr(1,2,3,4) - \overleftrightarrow\Tr(1,2,4,3) 
\bigr] \, \frac{n_s}{s} 
+ \bigl[ \overleftrightarrow\Tr(1,2,3,4) - \overleftrightarrow\Tr(1,3,2,4)
\bigr] \, \frac{n_t}{t} \nnl
&&\hskip0.0cm\null
+ \bigl[ \overleftrightarrow\Tr(1,3,2,4) - \overleftrightarrow\Tr(1,2,4,3)
\bigr] \, \frac{n_u}{u} \,.
\eea
Finally, one can convert the combination of traces
into structure constants using\footnote{%
This method is the guiding idea behind establishing a dual
representation of amplitudes in ref.~\cite{BernColorTraces}.}:
\be
\overleftrightarrow\Tr(1,2,3,4) - \overleftrightarrow\Tr(1,2,4,3)
\ =\ \tf^{a_1a_2e}\tf^{ea_3a_4}.
\ee
Thus, the full color-dressed four-point tree amplitude can be represented as 
\be
\CA_4(a_1,a_2,a_3,a_4)
\ =\ \frac{\tf^{a_1a_2e}\tf^{ea_3a_4}\,n_s}{s}
   + \frac{\tf^{a_2a_3e}\tf^{ea_4a_1}\,n_t}{t}
   + \frac{\tf^{a_1a_3e}\tf^{ea_2a_4}\,n_u}{u}
\ =\ \sum_{g\in\Gamma_4}\frac{c_g \, n_g}{p_g^2} \,,
\label{final4ptBCJ}
\ee
where $\Gamma_4$ is the set of cubic tree graphs in \fig{fig:stugraphs};
$c_g$ and $n_g$ are mappings providing, respectively, the color and
numerator factors of the graph $g$; and $1/p_g^2$ denotes the scalar 
propagator of the graph. Color-kinematics duality for four-point amplitudes
was first identified in refs.~\cite{FourPtBCJ}.

For a general $m$-point scattering amplitude at tree level,
the correspondence between color and kinematics~\cite{BCJ}
begins with the cubic-graph representation,
\be\label{cubicGaugeExpr} 
{\cal A}_m\ =\ \sum_{g \in \Gamma_m}
 \frac{c_g \, n_g}{\prod_{l \in P_g}{p_l^2}} \,\,, 
\ee
where the sum runs over all $m$-point cubic tree graphs $\Gamma_m$.
The product in the denominator runs over the set $P_g$ of all scalar
propagators $l$, with momenta $p_l$, that appear in the graph $g$.
In addition, we have omitted the Yang-Mills coupling constant for convenience.

Color-kinematics duality holds if the kinematic factors obey two criteria:
\begin{enumerate}
\item For a pair of graphs $g$ and $\hat{g}$ that are related by an
exchange of two external legs connected by a three-vertex,
the color factors are related by the antisymmetry of the structure
constant, $c_g = -c_{\hat{g}}$.  The numerator factors must also
be antisymmetric,
\be 
n_g = -n_{\hat{g}} \,.
\label{vertexFlip}
\ee
\item For a triplet of graphs $i,j,k$ that differ only by the
routing of lines in a four-point subgraph, as in \fig{fig:stugraphs},
the color Jacobi relations imply that $c_i + c_j + c_k = 0$.
Then the numerators must obey
\be
 n_i + n_j + n_k = 0 \,.
\label{Jacobi}
\ee
There are (coupled) sign ambiguities in defining the $c_g$ and $n_g$; hence
other presentations are possible, such as $c_i = c_j + c_k$ accompanied
by $n_i = n_j + n_k$.
\end{enumerate}
A set of numerators $n_g$ satisfying \eqns{vertexFlip}{Jacobi} is called a
color-kinematic dual representation. 

In the four-point example~(\ref{final4ptBCJ}), the one Jacobi
constraint on the numerators is
\be
n_s = n_t + n_u \,,
\label{4ptJac}
\ee
corresponding to the Jacobi relation
\be
\tf^{a_1a_2e}\tf^{ea_3a_4}
 = \tf^{a_2a_3e} \tf^{ea_4a_1}
 + \tf^{a_1a_3e} \tf^{ea_2a_4}.
\ee
In the general case, the constraints~(\ref{Jacobi}) are very powerful,
considerably reducing the number of independent color-ordered subamplitudes.
Naively, there are $(m-1)!$ independent subamplitudes
in \eqn{colordecomposition}, corresponding to the number of cyclically
inequivalent traces.  This number is reduced to $(m-2)!$ by
the group-theoretic Kleiss-Kuijf relations~\cite{KleissKuijf,DDDM}.
The linear relations that follow from consistency with eqs.~(\ref{Jacobi})
are known as the BCJ amplitude
relations~\cite{BCJ,MonodromyNBI1,MonodromyStieberger,MonodromyNBI2}.
They further reduce the number of independent subamplitudes to $(m-3)!$.

At the four-point level, there is only one BCJ relation,
\begin{equation}
  0 = s \, A_4(2,1,3,4) - t \, A_4(2,3,1,4).
\end{equation}
It reduces the two Kleiss-Kuijf independent subamplitudes down to a
single independent subamplitude. 

Although the existence of BCJ relations is intimately tied to the existence
of color-kinematic dual representations, there is no proof of equivalence
for generic theories including nonrenormalizable operators. 
A color-kinematic dual representation does imply the existence of BCJ 
amplitude relations.  Concerning the reverse direction,
explicit (albeit non-local) representations of dual
numerators $n_g$ in terms of color-ordered amplitudes have been
obtained for any number of external legs in (renormalizable) gauge theory,
based on string theory arguments~\cite{ExplicitForms}.
For MHV amplitudes, a construction of dual numerators has been given
based on the existence of a diffeomorphism Lie algebra for
the self-dual theory~\cite{MO}.
Below, we will test for color-kinematic dual representations of $F^3$
amplitudes by explicitly constructing or calculating the numerators.
At the same time we will also test the BCJ amplitude relations explicitly,
and provide a general argument for their validity.  These results further
support the equivalence of BCJ relations and color-kinematic dual
representations, at least in the $F^3$ case.

\subsection{Kawai-Lewellen-Tye relations}

Tree-level amplitudes in closed and open string theories are linked by the KLT
relations~\cite{KLT}, which arise from the fact that any closed-string vertex
operator can be represented as a product of two open-string vertex operators.
Closed-string scattering amplitudes are sums of products of left-moving
and right-moving open-string scattering amplitudes, multiplied by
various sine functions of the kinematics arising from contour deformations.
When one takes the low-energy, or infinite string tension 
($\alpha'\rightarrow 0$), limit of the KLT relations, the sine functions
collapse to Mandelstam variables, and the relations express gravity
amplitudes $M_m$ in terms of color-ordered amplitudes $A_m$ in gauge theory.
For three to six external legs, the field-theory KLT relations are given by,
\bea
M_3(1,2,3) &=& i\,A_3(1,2,3)\,\tA_3(1,2,3) \,, \nnl
M_4 (1,2,3,4) &=& - is_{12}\,A_4(1,2,3,4)\,\tA_4(1,2,4,3) \,, \nnl
M_5(1,2,3,4,5) &=& i s_{12}s_{34}\,A_5(1,2,3,4,5)\,\tA_5(2,1,4,3,5)
\ +\ \CP(2,3) , \nnl
M_6(1,2,3,4,5,6)&=& - i s_{12}s_{45}\,A_6(1,2,3,4,5,6)\nnl
&&\hskip0.2cm\null \times \Bigl[ s_{35}\,\tA_6(2,1,5,3,4,6)
+ (s_{34}+s_{35}) \, \tA_6(2,1,5,4,3,6) \Bigr]\nnl
&&\hskip0.2cm\null +\ \CP(2,3,4) \,,
\label{FTKLT}
\eea
where $s_{ij} = (k_i+k_j)^2$.
Formulae for higher-point amplitudes can be derived
straightforwardly~\cite{KLT}. In the above equation, `$+\CP$' indicates a sum
over all $p!$ nontrivial permutations of the $p$
indicated indices, acting on the term shown.  We refer to the amplitudes
$A_m$ ($\tA_m$) as coming from a `left-moving' (`right-moving') gauge theory.
In general, we may write the all-multiplicity KLT relations in the field
theory limit as a bilinear functional of the gauge-theory amplitudes,
\be
 M_m = {\rm KLT}_{{\rm FT},m}[ A_m, \tA_m ] \,.
\label{allmFTKLT}
\ee
The definitions of ${\rm KLT}_{{\rm FT},m}$ for $m=3,4,5,6$ follow
from eqs.~(\ref{FTKLT}).  For general $m$, the definitions follow
from the explicit formulae available in ref.~\cite{BernGaugeGravity}.
Below, we will use the notation in \eqn{allmFTKLT} in order to describe
extensions of the field-theory KLT relations to cover amplitudes generated
by higher-dimension operators.

\subsection{Double-copy construction of gravity amplitudes}

The field-theory limit of the KLT relations, combined with the
unitarity method~\cite{UnitarityMethod}, has provided, until recently, the most
important technical tool for perturbative calculations in maximal $\CN=8$
supergravity.  Multi-loop $\CN=8$ supergravity
integrands can be constructed by matching them against generalized
unitarity cuts, which are computed using products of $\CN=8$ supergravity
tree amplitudes.  The tree-amplitude products can be evaluated efficiently
using the KLT relations, together with the generalized cuts of
corresponding multi-loop amplitudes in $\CN=4$
super-Yang-Mills theory~\cite{BDDPR,ThreeloopNeq8,FourloopNeq8}.
However, this method still requires some effort, namely to process the
generalized cut information into a local gravity integrand.

In contrast, once an explicit color-kinematic dual representation is
found for a certain scattering amplitude in gauge theory, the double-copy
construction~\cite{BCJ,LoopBCJ,LagrangeBCJ} dramatically simplifies
the construction of the corresponding amplitude in the gravity theory.
According to this construction, one simply replaces the color factors
$c_g$ in \eqn{cubicGaugeExpr} with a second set of gauge-theory kinematical
numerators $\tilde{n}_g$, while keeping the remainder of the expression
untouched.  (The $\tilde{n}_g$ numerators come from the same right-moving gauge
theory discussed earlier in the context of the KLT relations.)
This method works at both the tree and loop
level~\cite{BCJ,LagrangeBCJ}.  The tree-level version, starting from
\eqn{cubicGaugeExpr}, is
\be\label{cubicGravExpr} 
  M_m\ =\ i \sum_{g \in \Gamma_m}
 \frac{n_g \, \tilde{n}_g}{\prod_{l \in P_g}{p_l^2}} \,\,.
\ee
For convenience, we have omitted overall factors of $(\kappa/2)$,
where $\kappa$ is the gravitational coupling.

While the KLT relations pertain exclusively to tree amplitudes,
the BCJ-inspired double-copy construction has been conjectured to
hold to all loop orders.  It has been
demonstrated to work in many examples~\cite{BBJ,loop5pt}. Indeed,
it has been used to obtain three- and four-loop integrands
in $\CN=8$ supergravity that make manifest, term by term, the correct
ultraviolet behavior of the full amplitude~\cite{LoopBCJ,FourLoopBCJ}.

In the next section, we will provide strong evidence that the
gauge-theory amplitudes produced by adding the operator $F^3$ to the
action also have a color-kinematic
dual representation.  In the section after that,
we will use this property to generate a double-copy construction of the
gravitational amplitudes produced by the operators $e^{-2\phi} R^2$ 
and $R^3$ in the effective action of the closed bosonic string.

\section{$F^3$ modification of gauge theory}\label{sec:F3}

In this section we consider the following addition to the YM action,
\be\label{F3}
O_\Lambda=\frac{1}{\Lambda^2}\Tr(F_\m^{~\n} F_\n^{~\r} F_\r^{~\m}) \,.
\ee
This operator, referred to as $F^3$, is the unique, gauge-invariant,
CP-even dimension-six operator built exclusively from gluon fields. The
prefactor $1/\Lambda^2$, where $\Lambda$ has mass-dimension one,
renders the operator four-dimensional. 

The antisymmetry of the gauge field strength, $F_{\r\m}=-F_{\m\r}$, 
enables us to immediately rewrite the color trace as
\be
\Tr(F_\m^{~\n} F_\n^{~\r} F_\r^{~\m})
\ =\ \sfrac{1}{2} \Tr([T^a,T^b]\,T^c) \, F_\m^{a\,\n} F_\n^{b\,\r} F_\r^{c\,\m}
\ =\ \sfrac{1}{2} \tf^{abc} \, F_\m^{a\,\n} F_\n^{b\,\r} F_\r^{c\,\m}\,.
\label{totalantisymF3}
\ee
Because the symmetries of the kinematic part of this operator
project onto the totally antisymmetric part of the trace,
the resulting $\tf^{abc}$ color structure for the three-vertex
generated by $F^3$ is the same as in usual YM theory. 

\begin{figure}
  \includegraphics[width=11cm]{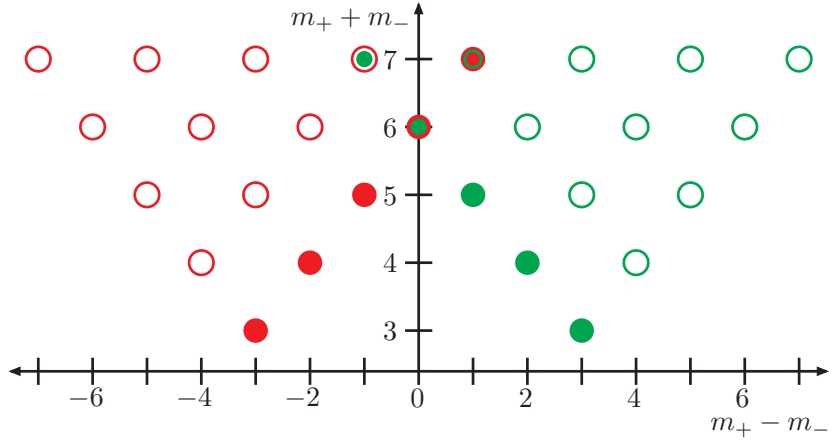}
  \caption{The $m$-gluon amplitudes produced by the operator $F^3$, where $m_+$
  and $m_-$ are the numbers of positive- and negative-helicity gluons,
  respectively. The left tower (red circles) of amplitudes is the self-dual
  sector, produced by $F^3_{\sst SD}$.  The right tower (green circles) is
  produced by the anti-self-dual operator $F^3_{\sst ASD}$. 
  For \MHV{} (\MHVbar) amplitudes with exactly three negative (positive)
  helicities, the circles are filled.  The figure is from
  ref.~\cite{DixonGloverKhoze}.}
  \label{fig:F3amps}
\end{figure}

In this article we will mostly consider amplitudes whose deviation from
the pure YM action is linear in $1/\Lambda^2$.  These amplitudes arise from one
insertion of the $F^3$ vertex, combined with any number of tree-level YM
interactions. The only exception occurs in~\sect{sec:Beyond}, where we
need to consider a diagram with two insertions of $F^3$ vertices, in order
to explain the difference between bosonic string theory and superstring
theory at second order in the inverse string tension. 

One can decompose the $F^3$ insertion into a
holomorphic (self-dual) and an antiholomorphic (anti-self-dual) part,
\bea
  O_\Lambda&=& O_{\Lambda_+} + O_{\Lambda_-} \,, \label{dualsplit}\\
  O_{\Lambda_+} &=& \frac{1}{\Lambda^2} \Tr(F_{{\sst SD}\,\m}^{~\n} 
F_{{\sst SD}\,\n}^{~\r} F_{{\sst SD}\,\r}^{~\m})
\, \equiv \, F^3_{\sst SD} \,, \nnl
  O_{\Lambda_-} &=&  \frac{1}{\Lambda^2} \Tr(F_{{\sst ASD}\,\m}^{~\n}
   F_{{\sst ASD}\,\n}^{~\r} F_{{\sst ASD}\,\r}^{~\m})
\, \equiv \, F^3_{\sst ASD} \,,
\eea
where
\be
F_{\sst SD}^{\mu\nu} = \hf(F^{\mu\nu}+\tilde{F}^{\mu\nu}) \ , \quad
F_{\sst ASD}^{\mu\nu} = \hf(F^{\mu\nu}-\tilde{F}^{\mu\nu}) \ , \quad
\tilde{F}^{\mu\nu} \equiv \ihf \epsilon^{\mu\nu\rho\sigma} F_{\rho\sigma} \ .
\label{SDASDdef}
\ee
This decomposition exposes the MHV structure of the amplitudes
produced by $F^3$~\cite{DixonGloverKhoze}.  Parity exchanges the two
operators, $O_{\Lambda_+} \leftrightarrow O_{\Lambda_-}$.

Combining one vertex from $F^3_{\sst SD}$ with an arbitrary number of pure YM
vertices leads to all the amplitudes in the left tower in \fig{fig:F3amps} (red
dots).  All amplitudes from the right tower (green dots), corresponding to
amplitudes originating from the anti-self-dual operator $F^3_{\sst ASD}$, can
be obtained immediately from the left tower using parity.  Parity exchanges
positive- and negative-helicity gluons, reflecting across the vertical axis in
the figure, and also complex conjugates all spinors.  The amplitudes produced
by $F^3$ in the overlapping region can be obtained by adding the self-dual and
anti-self-dual contributions.  We denote the amplitudes produced by $F^3$,
$F_{\sst SD}^3$ and $F_{\sst ASD}^3$, respectively, by $A_m^{F}$, $A_m^{F+}$
and $A_m^{F-}$, with $A_m^{F} = A_m^{F+} + A_m^{F-}$.

From now on, unless otherwise specified, we consider the left tower,
namely the amplitudes $A_m^{F+}$ produced by $F_{\sst SD}^3$.  Every
nonvanishing amplitude of this type should have at least three
negative-helicity gluons~\cite{DixonGloverKhoze}.  That is,
\be
A_m^{F+}(1^+,\ldots,m^+)\ =\ A_m^{F+}(1^+,\ldots,j^-,\ldots,m^+)
\ =\ A_m^{F+}(1^+,\ldots,j^-,\ldots,k^-,\ldots,m^+)\ =\ 0,
\ee
where $j$ and $k$ are the only negative-helicity gluons.
Amplitudes with exactly three negative-helicity gluons and an
arbitrary number of positive-helicity gluons are called MHV
amplitudes.  (In the right tower generated by $F_{\sst ASD}^3$,
the amplitudes with exactly three positive-helicity gluons, and the
rest negative, form the $\overline{\rm MHV}$ class.)
The lowest interaction derivable from the self-dual part
belongs to the MHV class and is a three-point vertex with three negative
helicities,
\be
A_3^{F+}(1^-,2^-,3^-)\ =\ i \, \spa1.2\spa2.3\spa3.1 \,.
\label{F3mmm}
\ee
It is not an on-shell amplitude for real momenta, but
it is perfectly well-defined and nonzero for complex momenta.
In writing the amplitudes $A_m^{F}$, $A_m^{F+}$ and $A_m^{F-}$, we
omit for convenience an overall factor of $-3g^{m-2}/\Lambda^2$.


\subsection{Known $F^3$ amplitudes}

The Parke-Taylor formula~\cite{ParkeTaylor} for MHV tree amplitudes
in YM theory reads,
\be
A_m(1^+,\ldots,j^-,\ldots,k^-,\ldots,m^+)
\ =\ i \, \frac{ {\spa{j}.k}^4}{\spa1.2\spa2.3\cdots\spa{m}.1} \,,
\label{PTformula}
\ee
where $j$ and $k$ are the only negative-helicity gluons, and
the usual spinor-helicity brackets $\spa{j}.k$ have been used.
Analogous expressions for the MHV amplitudes in the $F^3$ sector,
$A_m^{F+}$, were first obtained for $m=4,5$~\cite{DixonShadmi},
and were later generalized to other values of $m$~\cite{DixonGloverKhoze}.
They take the form,
\be
  A_m^{F+}(1^+,\ldots,j^-,\ldots,k^-,\ldots,l^-,\ldots,m^+)\ =\ 
\ i\, \frac{ {\spa{j}.k}^2 {\spa{k}.l}^2 {\spa{l}.j}^2 }
     {\spa1.2\spa2.3\cdots\spa{m}.1} \,,
\label{F3MHV}
\ee
where $j$, $k$ and $l$ are the only negative-helicity gluons.

Besides the MHV amplitudes in \eqn{F3MHV}, several further four-parton
amplitudes have been computed~\cite{DixonShadmi,DixonGloverKhoze}. 
At the four-point level (and three-point level) the $F^3$ amplitudes are
orthogonal to the pure YM amplitudes.  The only non-vanishing, non-MHV
four-point amplitude produced by $F^3_{\sst SD}$ is
\be
A_4^{F+}(1^-,2^-,3^-,4^-)\ =\ \frac{-2 \, i \, s_{12}s_{23}s_{13}}
  {\spb1.2\spb2.3\spb3.4\spb4.1} \,,
\label{F3mmmm}
\ee
which has a very simple, cyclicly symmetric form.

\subsection{CSW rules for $F^3$ amplitudes}

In order to generate further $F^3$ amplitudes, beyond those given above, 
we adapt the CSW formalism~\cite{CSW} based on MHV vertices to
the present context~\cite{DixonGloverKhoze, BoelsF3}.
An N${}^k$MHV amplitude $A_m^{F+}$ is generated by precisely one
$F_{\sst SD}^3$ MHV vertex (the off-shell continuation of \eqn{F3MHV})
and $k$ YM-theory MHV vertices (the off-shell continuation
of the Parke-Taylor formula~(\ref{PTformula})).
Because the $F_{\sst SD}^3$ MHV vertex provides three negative helicities,
not two, an N${}^k$MHV amplitude contains $k+3$ negative-helicity gluons.
The prescription for continuing off shell a holomorphic (right-handed)
spinor $\la_a$ that enters a CSW propagator is exactly as in YM theory.
Namely, one introduces a reference spinor $\xi$ via the replacement
\bea
  (\la_a)_\a\ \rightarrow\ (p_a)_{\a\ad}\xi^\ad \,,
  \label{CSWoffshell}
\eea
whenever leg $a$ is off shell.

\begin{figure}
  \includegraphics[width=13cm]{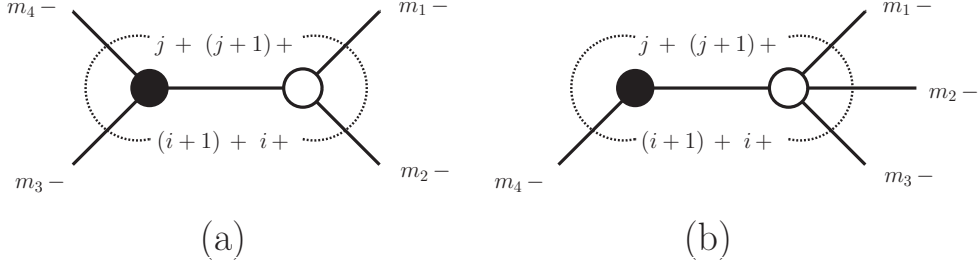}
  \caption{Next-to-MHV $F^3$ amplitudes are constructed by sewing
 together two MHV vertices, one for $F^3$ (white circle) and one for
 ordinary gauge theory (black dot).  There are two distinct ways of
 distributing the four negative-helicity gluons, (a) and (b).
 The dotted lines indicate that an arbitrary number of positive-helicity
 gluons may be present. }
  \label{fig:F3nmhv}
\end{figure}

At next-to-MHV (NMHV), the two distinct classes of MHV-vertex graphs are
shown in \fig{fig:F3nmhv}.  Following refs.~\cite{GGK,DixonGloverKhoze},
they generate the amplitudes
\be
A_m^{F+}(m_1^-,m_2^-,m_3^-,m_4^-)\ =\ i\, 
\frac{\sum_{C(m_1,m_2,m_3,m_4)} \Bigl[ A_m^{(a)}(m_1,m_2,m_3,m_4)
+ A_m^{(b)}(m_1,m_2,m_3,m_4) \Bigr]}{\prod_{l=1}^m \spa{l,}.{l+1}} 
   \,,
\label{F3NMHVgen}
\ee
where we have omitted the labels of the $(m-4)$ positive-helicity gluons.
The sum over $C(m_1,m_2,m_3,m_4)$ is over the four cyclic permutations
of the four negative-helicity labels.  The two distinct classes of
graphs evaluate to
\bea
A_m^{(a)}(m_1,m_2,m_3,m_4) &=& \sum_{i=m_2}^{m_3-1} \sum_{j=m_4}^{m_1-1}
 \frac{ {\spa{m_1}.{m_2}}^2 
  \langle m_1^- | \slash\!\!\! q_{j+1,i} | \xi^-\rangle^2
  \langle m_2^- | \slash\!\!\! q_{j+1,i} | \xi^-\rangle^2
        {\spa{m_3}.{m_4}}^4 }
      { D(i,j,q_{j+1,i}) } \,,~~~~~~
\label{F3NMHVa}\\
A_m^{(b)}(m_1,m_2,m_3,m_4) &=& \sum_{i=m_3}^{m_4-1} \sum_{j=m_4}^{m_1-1}
 \frac{ {\spa{m_1}.{m_2}}^2 {\spa{m_2}.{m_3}}^2 {\spa{m_3}.{m_1}}^2 
       \langle m_4^- | \slash\!\!\! q_{j+1,i} | \xi^-\rangle^4 }
      { D(i,j,q_{j+1,i}) } \,, \label{F3NMHVb}
\eea
where $q_{j+1,i} = k_{j+1} + k_{j+2} + \cdots + k_i$, and all external
leg labels are defined modulo $m$.   The `dressed' propagator is,
\be
D(i,j,q) = \langle i^-| \slash\!\!\! q | \xi^-\rangle
         \langle(i+1)^-| \slash\!\!\! q | \xi^-\rangle
         \langle j^-| \slash\!\!\! q | \xi^-\rangle
         \langle(j+1)^-| \slash\!\!\! q | \xi^-\rangle
  \frac{q^2}{ \spa{i,}.{i+1} \spa{j,}.{j+1} } \,.
\label{dressedprop}
\ee
The sum over $i,j$ is over the ways of distributing the positive-helicity
gluons.  In \fig{fig:F3nmhv}(b) there must be at least one positive-helicity
gluon attached to the left vertex.  If there is not, the factor 
$\langle m_4^- | \slash\!\!\! q_{j+1,i} | \xi^-\rangle
= - \langle m_4^- | \ksl_{m_4} | \xi^-\rangle = 0$
kills the corresponding term in any event.
We have also implemented the general formula for N${}^2$MHV amplitudes,
which involves 14 distinct classes of CSW diagrams.  Finally, we worked out
the one N${}^3$MHV amplitude at the six-point level, $A_6^{F+}$ with
all negative helicities, in terms of 56 CSW diagrams.

Employing this modified CSW procedure, we have calculated the following
amplitudes:
\bea
  && A_5^{F+}(1^-,2^-,3^-,4^-,5^+) \,, \nnl
  && A_5^{F+}(1^-,2^-,3^-,4^-,5^-) \,, \nnl
  && A_6^{F+}(1^-,2^-,3^-,4^-,5^+,6^+) \,, \nnl
  && A_6^{F+}(1^-,2^-,3^-,4^+,5^-,6^+) \,, \nnl
  && A_6^{F+}(1^-,2^-,3^+,4^-,5^-,6^+) \,, \nnl
  && A_6^{F+}(1^-,2^-,3^-,4^-,5^-,6^+) \,, \nnl
  && A_6^{F+}(1^-,2^-,3^-,4^-,5^-,6^-) \,.
  \label{availableF3amps}
\eea
In addition, we obtained the generic NMHV (four-minus) and N${}^2$MHV
(five-minus) amplitudes with seven or more external legs.
The results are fairly
lengthy.  While the complexity of the expressions~(\ref{F3MHV}) for the
MHV amplitudes from $F^3$ is at the same level as that of the Parke-Taylor
amplitudes~(\ref{PTformula}), the forms we find in the higher NMHV sectors
become quite complex in comparison with the corresponding YM amplitudes.
Of course, they can still be evaluated numerically quite easily, so that one
can test, for example, the required independence of the expressions
from the reference vector $\xi$.  We also computed the $F^3$ amplitudes
numerically from Feynman diagrams through six external legs, using the 
{\sc MadGraph} package~\cite{MadGraph5,UFO,ALOHA}; the Feynman vertices for
the $F^3$ interaction were generated by {\sc FeynRules}~\cite{FeynRules}.
We found complete numerical agreement with the CSW expressions.
For the six-point amplitudes with three positive and three negative
helicities, the agreement is with the sum of the self-dual and
anti-self-dual contributions, $A_6^{F+} + A_6^{F-}$ (see \fig{fig:F3amps}).

One NMHV amplitude that can be simplified easily from the CSW
representation is the five-point case,
\be
A_5^{F+}(1^-,2^-,3^-,4^-,5^+) \, = \, i\, \biggl[ 
\frac{ {\spa1.2}^2 {\spb2.5}^2 \spb1.5 }
     { \spb1.2 \spb2.3 \spb3.4 \spb4.5 }
+ \frac{ {\spa3.4}^2 {\spb3.5}^2 \spb4.5 }
     { \spb1.2 \spb2.3 \spb3.4 \spb1.5 }
+ 2 \, \frac{ {\spa1.4}^2 \spa1.3 \spa2.4 }
            { \spb2.3 \spa4.5 \spa1.5 } \biggr] \,. \nnl
~\label{mmmmpSimple}
\ee
The N${}^2$MHV all-minus five-point amplitude is also simple, in part
because it is cyclicly symmetric,
\be
A_5^{F+}(1^-,2^-,3^-,4^-,5^-)\, =\, i\, 
\frac{ \sum_{i=1}^5 \bigl[ s_{i,i+1} s_{i+1,i+2} ( s_{i,i+2} - 2 s_{i+3,i+4} )
   + \varepsilon(1,2,3,4) \, s_{i,i+1} \bigr] } 
  { \spb1.2\spb2.3\spb3.4\spb4.5\spb5.1 } \,,
\label{mmmmmSimple}
\ee
where
$\varepsilon(i,j,l,m) \equiv
4i\varepsilon_{\mu\nu\sigma\rho} k_i^{\mu}k_j^{\nu}k_l^{\sigma}k_m^{\rho}$
is the contracted Levi-Civita tensor.

We also provide one NMHV example at the six-point level,
\be
A_6^{F+}(1^-,2^-,3^+,4^-,5^-,6^+)\ =\ \ i\,\Bigl[ 
X_6 + X_6|_{1\leftrightarrow5,\,2\leftrightarrow4} \Bigr] \,,
\label{mmpmmp}
\ee
where
\bea
X_6 &=& 
\frac{ {\spb3.6}^3  {\spa1.4}^2 {\spa1.5}^2 \spa4.5 }
             { \spb2.3 \spa5.6 \spa6.1 \, \spb2.6 
               \, \spab1.{(2+3)}.6 \spab4.{(2+3)}.6 }
\label{mmpmmpA}\\
&& \hskip0cm\null
+  \frac{ \spa1.2 \spaa1.{(4+5)}.{(4+6)}.5 }
   { \spa2.3 \spb4.5 \spb4.6 s_{56} \spa6.1 s_{123}
   \, \spab1.{(2+3)}.6 \spab3.{(1+2)}.6 } \nnl
&& \hskip1.2cm\null
 \times \biggl[ \Bigl( \spa1.2 \spb6.4 \spa4.5 \spb5.6 \Bigr)^2
             + \Bigl( \spab1.{(4+5)}.6 \spab2.{(4+5)}.6 \Bigr)^2 \biggr] \nnl
&& \hskip0cm\null
+ \frac{1}{2} \frac{ {\spa1.5}^2  {\spa2.4}^2  \bigl[ 
      ( \spa2.4 \spb6.1 \spa1.5 \spb5.6 )^2
    + ( \spab2.{(3+4)}.6 \spab4.{(2+3)}.6 )^2 \bigr] }
   { \spa2.3 \spa3.4 s_{56} s_{61} s_{234} 
    \, \spab2.{(3+4)}.6 \spab4.{(2+3)}.6 } \,, \nonumber
\eea
and $s_{ijl} = (k_i+k_j+k_l)^2$ is a multi-particle invariant.

Finally, the N${}^3$MHV all-minus six-point amplitude is cyclicly
symmetric and lacks multi-particle poles, allowing us to simplify
it as well,
\be
A_6^{F+}(1^-,2^-,3^-,4^-,5^-,6^-)\ =\ i\,  
\frac{ \sum_{i=1}^6 f_i^{6-} }
  { \spb1.2\spb2.3\spb3.4\spb4.5\spb5.6\spb6.1 } \,,
\label{mmmmmmSimple}
\ee
where
\bea
f_i^{6-} &=& s_{i,i+1}^2 ( s_{i-1,i} + s_{i+1,i+2} - s_{i+3,i+4} )
\, + \,\frac{1}{2} s_{i,i+1,i+2} \bigl[
    s_{i-2,i-1,i} \, s_{i-1,i,i+1} + s_{i-1,i,i+1} \, s_{i,i+1,i+2}
\nnl &&\hskip1.0cm\null 
   + s_{i,i+1,i+2} \, s_{i-2,i-1,i}
   + 2 \, s_{i+1,i+2} \, ( s_{i-2,i-1,i} + s_{i,i+1} )
\nnl &&\hskip1.0cm\null
   - s_{i,i+1} \, s_{i+3,i+4}
   - s_{i+1,i+2} \, s_{i+4,i+5}
   - 3 \, s_{i+2,i+3} \, s_{i-1,i} \bigr]
\nnl &&\hskip0.0cm\null
   - \frac{1}{2} \, \varepsilon(i,i+1,i+2,i+3) \, \bigl[
    s_{i-2,i-1,i} + s_{i-1,i,i+1} + s_{i,i+1,i+2} + 2 \, s_{i+1,i+2} \bigr] \,.
\label{f6minus}
\eea

In view of the similarity between the two sets of MHV amplitudes, YM theory and
$F^3$, it would be very interesting to search for alternative, manifestly
$\xi$-independent, all-multiplicity forms of the $F^3$ amplitudes, analogous to
those provided by the superconformal $R$ invariants in YM
theory~\cite{HennTreeAmps}. In addition, one could imagine that --- although
the theory is not supersymmetrizable --- there might be a compact
representation in terms of momentum twistors, similar to the one for
pure YM theory~\cite{MomTwistor}.

\subsection{Color-kinematics duality for $F^3$ amplitudes}

We now consider color-kinematics duality for $F^3$ amplitudes with
an arbitrary number of legs.  Just as in the YM case~\cite{BCJ}, we
generalize the equations from the four-point example in
subsection~\ref{ssec:BCJrep}.  A color-kinematic dual description of an
amplitude exists, if one can find a set of kinematic numerators
$n_g$ that satisfies three sets of equations:
\begin{itemize}
  \item \textit{automorphism equations} ensuring the total antisymmetry of
    the kinematic numerators at each vertex, as dictated by color, 
   {\it i.e.}~the antisymmetry of the structure constants
   (generalizations of \eqn{vertexFlip}), 
  \item \textit{Jacobi equations} ensuring that any numerators for
    diagrams whose color structures are related by Jacobi identities
    are related in the same way, via \eqn{Jacobi},
  \item \textit{amplitude expansion} --- the color-ordered amplitudes
    are dictated by the $n_g$, after expanding the color factors out into
    traces in the fundamental representation using \eqn{structureConstants}.
\end{itemize}
Fortunately, all of the above equations are linear, which makes it
straightforward to obtain solutions analytically and numerically. However,
given the complexity of the expressions obtained for N${}^k$MHV amplitudes in
the last subsection, it is often more convenient to test numerically whether
the set of all equations allows for a solution at all.  As discussed at the end
of subsection~\ref{ssec:BCJrep}, it should be equivalent
(and is often simpler) to test
directly the linear BCJ relations between color-ordered amplitudes~\cite{BCJ}.
For example, at the five- and six-point levels we test for the vanishing of
\bea
&&s_{12} s_{45} A_5^{F+}(1,2,3,4,5)
+ s_{13} s_{24} A_5^{F+}(1,3,4,2,5)
- s_{14} (s_{24}+s_{25}) A_5^{F+}(1,4,3,2,5) \,, \nn\\
\label{BCJtest5}\\
&&( s_{14} + s_{45} + s_{46} ) A_6^{F+}(1,2,3,4,5,6)
+ ( s_{14} + s_{46} ) A_6^{F+}(1,2,3,5,4,6)
\nn\\
&&\hskip0cm\null
+ s_{14} A_6^{F+}(1,2,3,5,6,4)
- s_{24} A_6^{F+}(1,2,4,3,5,6) \,,
\label{BCJtest6}
\eea
plus permutations of these equations.

It is amusing to note that consistency with the BCJ relations determines
certain contributions that cannot be determined using factorization
limits with real momentum.  For example, in \eqn{mmmmmSimple} the term
proportional to $\varepsilon(1,2,3,4)$
vanishes in all (real) collinear limits.  Its coefficient is fixed,
however, by imposing the vanishing of \eqn{BCJtest5}.

\subsection{Recycling of numerators}
\label{Recycling}

The close similarity between \eqns{PTformula}{F3MHV} was one motivation to
explore the consistency of $F^3$ amplitudes with color-kinematics duality.
Indeed, as will be pointed out below, at the MHV level the duality
for $F^3$ amplitudes is already manifest, given a dual representation
for YM theory.

The MHV amplitudes~(\ref{PTformula}) in pure YM theory can be
split into two pieces.  The state-dependent part ${\spa{p}.q}^4$,
where $p,q$ are the two negative-helicity gluons, is determined by 
a Grassmann integral in an on-shell superspace~\cite{Nair}.
The state-independent remnant,
\be
A_m^{\rm MHV,rem} \equiv \frac{i}{\spa1.2\spa2.3\cdots\spa{m}.1} \,,
\label{MHVdressingRemnant}
\ee
carries all of the dependence on the color ordering.
The BCJ relations for color-ordered amplitudes relate amplitudes
with the same external states but different color ordering.
Therefore at the MHV level they are equivalent to relations among
the remnants.  We define
\be
 n_g^{\rm MHV,rem} =  \frac{n_g^{{\rm YM},p,q}}{{\spa{p}.{q}}^4}
\label{remnumerator}
\ee
as the set of BCJ numerator factors out of which the state-independent
remnant is built.  For example, at the four-point level, we have
\be
\frac{i}{\<12\>\<23\>\<34\>\<41\>}
= \frac{n_s^{\rm MHV,rem}}{s} + \frac{n_t^{\rm MHV,rem}}{t} \,.
\ee

Given the existence of duality-satisfying
numerators $n_g^{{\rm YM},p,q}$ for MHV
amplitudes of arbitrary multiplicity in YM theory, we can easily obtain
dual numerators for other amplitudes, provided that they allow a similar
splitting, into a factor that is independent of the color ordering, multiplied
by the {\it same} remnant factor as in \eqn{MHVdressingRemnant}.  The prime
example of amplitudes obeying this criterion are the MHV amplitudes from the
$F^3_{\sst SD}$ sector. They split according to  
\be
A_m^{F_+}(\ldots,j^-,\ldots,k^-,\ldots,l^-,\ldots)
= \<jk\>^2\<kl\>^2\<lj\>^2 \times \frac{i}{\<12\>\<23\>\cdots\<m1\>}\,.
\label{MHVF3dressingRemnant}
\ee
Thus the numerator for an $F^3$ MHV amplitude can be obtained trivially
from the numerator for a pure YM MHV amplitude via
\be
n_g^{F^+,j,k,l}
= \<jk\>^2\<kl\>^2\<lj\>^2 \ n_g^{\rm MHV,rem}
=\ \frac{ {\spa{j}.k}^2 {\spa{k}.l}^2 {\spa{l}.j}^2 }
                  { {\spa{p}.q}^4 }  \ n_g^{{\rm YM},p,q} \,,
\label{MHVF3fromYM}
\ee
where $j,k,l$ and $p,q$ denote the negative-helicity gluons in the $F^+$ and YM
amplitudes respectively. 

Another example is provided by the all-minus four-point $F^3$
amplitude~(\ref{F3mmmm}). Here we use the splitting,
\be
A_4^{F^+}(1^-,2^-,3^-,4^-) = -2\,stu \times \frac{i}{[12][23][34][41]} \,,
\label{F3mmmmsplit}
\ee
where the remnant can be identified as the parity conjugate of the 
four-point MHV
remnant in \eqn{MHVdressingRemnant}.  This splitting works because $stu$ is
symmetric under all permutations, and hence independent of the color ordering.
Therefore a solution can be obtained from a solution for a pure YM four-point
amplitude via  
\be
\label{recyclemmmm}
n_g^{F^+,(----)}
= -2 \, stu \, \Bigl[ n_g^{\rm MHV,rem} \Bigr]^\dagger
= -2 \, \frac{stu}{[pq]^4}\, \Bigl[ n_g^{{\rm YM},p,q} \Bigr]^\dagger \,,
\ee
where the dagger denotes the spinor conjugation operation (parity).

Of course, one could have arrived at the same result in the conventional way
described in the last subsection. For the four-point amplitude, the cubic tree
graphs are universal and thus all contributions to $A_4^{F^+}(1^-,2^-,3^-,4^-)$
and permutations are shown in \fig{fig:stugraphs}.  While the automorphism
equations~(\ref{CubicAntisymmetry}) and the Jacobi identities~(\ref{4ptJac})
thus agree with pure YM theory, the amplitude equations are now supposed to
yield the appropriate $F^3$ amplitudes, for example,
\be
  A_4^{F+}(1^-,2^-,3^-,4^-)\ =\ \frac{n_s}{s}+\frac{n_t}{t}.
\ee
The resulting set of equations determining the numerators does not
have a unique solution.  Instead, it exhibits the same one-parameter
generalized gauge freedom as explored for pure YM theory in ref.~\cite{BCJ}.

This observation does not come as a surprise; it follows from the existence
of the recycled numerators~(\ref{recyclemmmm}) for this particular amplitude.
The YM remnant numerators possess the generalized gauge freedom, and this
property carries over to the recycled numerator via \eqn{recyclemmmm}.

A particularly nice numerator solution, reproducing \eqn{F3mmmm}, reads
\bea
  n_s&=&-\frac{2\, i}{3}\frac{s^2t(u-t)}{[12][23][34][41]} \,, \nnl
  n_t&=&-\frac{2\, i}{3}\frac{st^2(u-s)}{[23][34][41][12]} \,, \nnl
  n_u&=&-\frac{2\, i}{3}\frac{tu^2(s-t)}{[14][42][23][31]} \,,
  \label{solution4ptmmmm}
\eea
where the generalized gauge freedom has been used to find a
particularly symmetric expression. In fact, this is one of the
situations where one can find a so-called \textit{symmetric
numerator}, that is, a function assigning every graph with a
particular labeling a valid BCJ numerator. The following expression
reproduces~\eqn{solution4ptmmmm} for $n_s=n[1,2,3,4]$,
$n_t=n[2,3,4,1]$ and $n_u=[3,1,4,2]$:
\bea
  n[a,b,c,d]=-\frac{2\, i}{3}
\frac{s_{ab}s_{bc}}{[ab][bc][cd][da]}s_{ab}(s_{ac}-s_{bc})\,.
  \label{symmetricF3}
\eea
Symmetric numerators for five- and six-point MHV YM amplitudes have
been identified and discussed in ref.~\cite{Virtuous}.

\subsection{Monodromy relations and $F^3$ amplitudes}

At the end of subsection~\ref{ssec:BCJrep} we discussed the likely
equivalence of the BCJ amplitude relations and the existence of
color-kinematic dual representations. Assuming this equivalence,
there is a very good argument in favor of the existence of color-kinematic
dual representations
for all $F^3$ amplitudes, which originates in the string-theory monodromy
relations~\cite{MonodromyNBI1,MonodromyStieberger,MonodromyNBI2}.
These relations connect open-string amplitudes with different cyclic orderings
of the external legs.  They can be derived by deforming the contours for
the integration of open-string vertex operators along the boundary of the
world-sheet~\cite{Plahte}.

The monodromy relations read,
\begin{eqnarray}
  A^{\rm string}(1,2,3,4,\ldots,m)&
\!+\!&e^{i\a'\pi s_{12}}A^{\rm string}(2,1,3,4,\ldots,m)
+e^{i\a'\pi (s_{12}+s_{13})} A^{\rm string}(2,3,1,4,\ldots,m)\nnl
    &&\hspace{-1cm}+\cdots + e^{i\a'\pi
      (s_{12}+s_{13}+\ldots+s_{1,m-1})}A^{\rm string}(2,3,4,\ldots,m-1,1,m)=0.
  \label{Monodromy1}
\end{eqnarray}
We now expand the exponential in \eqn{Monodromy1} in $\a'$, and also the 
string amplitude,
\begin{equation}
  A^{\rm string} = A^{\YM}+\a'A^{\a'}+\cdots\,,
\end{equation}
where the first correction $A^{\a'}$ is directly proportional to the 
$F^3$ amplitude, $A^{\a'} \propto A^F$.  Different terms in the expansion
of \eqn{Monodromy1} provide different information.  By choosing the
polarization vectors to be real (linear polarizations), we can consider
$A^{\YM}$ and $A^{\a'}$ to be real.  (Later we can take linear combinations
with complex coefficients to obtain helicity amplitudes.)

The $\CO(\a'^0)$ term in the expansion leads to the standard photon ($U(1)$)
decoupling identity for YM amplitudes,
\bea
\label{YMdecoupling}
0 &=& A^{\YM}(1,2,3,4,\ldots,m)+A^{\YM}(2,1,3,4,\ldots,m)\nnl
&&\hspace{0.5cm}+A^{\YM}(2,3,1,4,\ldots,m)
+\cdots +A^{\YM}(2,3,4,\ldots,m-1,1,m)\,,
\eea
while the imaginary part of the $\CO(\a')$ term yields the 
BCJ relations for YM amplitudes,
\bea
0 &=&
  s_{12}A^{\YM}(2,1,3,\ldots,m)
 + (s_{12}+s_{13})A^{\YM}(2,3,1,4,\ldots,m)\nnl
    &&\hspace{0.5cm}
+\cdots + (s_{12}+s_{13}+\ldots+s_{1,m-1})A^{\YM}(2,3,4,\ldots,m-1,1,m) \,.
\label{YMBCJ}
\eea
In contrast, the real part of the $\CO(\a')$ term and the imaginary
part of the $\CO(\a'^2)$ term yield, respectively, the 
photon-decoupling and BCJ relations for the $F^3$ amplitudes,
\bea
\CO(\a'^1):\quad 0&=&A^{\a'}(1,2,3,4,\ldots,m)+A^{\a'}(2,1,3,4,\ldots,m)\nnl
&&\hspace{0.5cm}+A^{\a'}(2,3,1,4,\ldots,m)
+\cdots + A^{\a'}(2,3,4,\ldots,m-1,1,m) \,,\label{Monodromy2a}\\
\CO(\a'^2):\quad 0&=&
s_{12}A^{\a'}(2,1,3,\ldots,m)
\!+\!(s_{12}+s_{13})A^{\a'}(2,3,1,4,\ldots,m)\nnl
    &&\hspace{0.5cm}
+\cdots + (s_{12}+s_{13}+\ldots+s_{1,m-1})A^{\a'}(2,3,4,\ldots,m-1,1,m)\,.~~~~
  \label{Monodromy2}
\end{eqnarray}
In other words, the BCJ relations for the $F^3$ amplitudes can be derived
simply by using the string-theory monodromy relations, and the fact that
$F^3$ is the unique operator appearing at $\CO(\a')$ in the low-energy
effective action of the open bosonic string.

We have tested the $F^3$ amplitudes for the availability of color-kinematic
dual numerators (satisfying antisymmetry~(\ref{vertexFlip}) and kinematic
Jacobi relations~(\ref{Jacobi})).  We have
found such a representation to be indeed present in all cases.  At the
four-point level, the representations can be exhibited analytically; the
only independent cases are the all-minus and MHV cases discussed above.
For the five- and six-point cases we generally relied on numerical checks.  

Interestingly, the representations for $m$-point amplitudes exhibit
the same generalized gauge freedom as the numerators for YM amplitudes
with the same number of external legs.  The dimension of the vector space
of numerator solutions parametrizing the generalized gauge freedom in
pure gauge theory is
\be
  \text{dim}_{\rm ggf}\ =\ (m-2)! \, - \, (m-3)! \,,
  \label{dimModuliSpace}
\ee
which is exactly the dimension of the solution space we find for
color-kinematic dual representations of $F^3$ amplitudes. 

Given the availability of the dual representations for amplitudes
produced by the $F^3$ modification of pure gauge theory, we expect to
be able to use the double-copy construction to generate amplitudes in
a theory of gravity.  After carrying out this procedure in the next
section, we then address the question: to which theory of gravity do these
amplitudes belong?

\section{Squaring to gravity}\label{sec:R3}

In the last section we established the BCJ relations and color-kinematic dual
representations for gauge-theory amplitudes with one insertion of a vertex 
originating from the $F^3$ operator.  In YM theory, the existence of a
dual representation is a sufficient criterion for the double-copy
construction of amplitudes in Einstein gravity and its various
supersymmetrizations, given that the amplitudes behave well
for large complex BCFW shifts~\cite{LagrangeBCJ}.
Here we consider the set of gravitational amplitudes obtained by
applying the double-copy construction to the $F^3$-deformed gauge-theory
amplitudes from the last section.  The amplitudes are not as well behaved
under large shifts as in the YM case.  On the other hand, we know there
is a gravity-gauge-theory relation of some type, because the $F^3$ operator
is part of the effective action for the open bosonic string.
We will compare the double-copy results with those derived from the
KLT relations and discuss the modified gravitational action from which 
they can be obtained.

\subsection{Set of amplitudes}

In order to obtain the double copy of the $F^3$ amplitudes, we insert two
copies of the numerators calculated in the last section into
\eqn{cubicGravExpr}. Let us first consider the class of amplitudes resulting
from combining numerators from two $F^+$-amplitudes, which will be
denoted\footnote{%
In contrast to the $+$ in $F^+$ introduced in~\sect{sec:F3},
the superscript $+$ in $R^+$ does not imply that the resulting amplitudes
are matrix elements of the self-dual part of some operator.  We will see below
that this is not the case.}
by a superscript $R^+$. 
Unless otherwise noted, we will always combine pairs of external gluons
$g^\pm$ with the same helicity in the YM amplitudes, so as to yield only
gravitons $G^\pm$ after double-copying:
\be
G^+ = g^+ \otimes g^+ \qquad \text{and} 
\qquad G^- = g^- \otimes g^-\,.
\ee
Choosing opposite helicities, $g^\pm \otimes g^\mp$, will generate
the dilaton and axion scalar states, as we will see later.

The basic building blocks for the set of gravitational amplitudes are the
three-point identical-helicity vertices, which are exactly the square
of their $F^3$ counterparts:
\be
  M_3^{R+}(1^-,2^-,3^-)\ =\ -i \, ( \spa1.2\spa2.3\spa3.1 )^2
  \ =\ i \, \bigl[ A_3^{F+}(1^-,2^-,3^-) \bigr]^2\,.
  \label{3ptKLT}
\ee
Using the notation introduced in \eqn{allmFTKLT}, we can write
\eqn{3ptKLT} as
\be
  M_3^{R+}(1^-,2^-,3^-)\ =\ 
{\rm KLT}_{{\rm FT},3}[ A_3^{F^+}(1^-,2^-,3^-), A_3^{F^+}(1^-,2^-,3^-) ]\,.
\label{3ptKLT2}
\ee  
For the four-graviton amplitude with three negative-helicity gravitons,
the double-copy construction yields
\bea
 M_4^{R+}(1^-,2^-,3^-,4^+) &=& 
  \frac{(\spa1.2\spa2.3\spa3.1)^4}{{\spa1.2}^8}
  \, M_4(1^-,2^-,3^+,4^+)
  \label{4ptR3MHVa}\\
&=& -i \, ( \spb4.1\spa1.3\spb3.4 )^2
\, \frac{\spa1.2\spa2.3\spa3.1}{\spb1.2\spb2.3\spb3.1} \,,
\label{4ptR3MHVb}
\eea
which is in agreement with
\be
{\rm KLT}_{{\rm FT},4}[ A_4^{F^+}(1^-,2^-,3^-,4^+),A_4^{F^+}(1^-,2^-,3^-,4^+) ]\,.  
\label{4ptR3MHVKLT}
\ee
In fact, the general form of the $m$-graviton $R^+$ amplitudes with exactly
three negative-helicity gravitons,
\bea
&&M_m^{R+}(1^+,\ldots,j^-,\ldots,k^-,\ldots,l^-,\ldots,m^+)\nnl
&&\hskip1.5cm 
=\ \frac{ ( {\spa{j}.k} {\spa{k}.l} {\spa{l}.j} )^4 }{ {\spa{p}.q}^8 }
  M_m(1^+,\ldots,p^-,\ldots,q^-,\ldots,m^+) \,,
\label{R3MHV}
\eea
follows straightforwardly from \eqn{MHVF3fromYM}.

Consider now the first amplitude with four negative-helicity gravitons.
One finds that the double-copy construction yields
\bea
  M_4^{R+}(1^-,2^-,3^-,4^-)
&=& 4 \frac{(stu)^2}{{\spb3.4}^8} \,  M_4(1^-,2^-,3^+,4^+)
\label{R3mmmma}\\
&=& 4 \, i \, stu 
\frac{\spa1.2\spa2.3\spa3.4\spa4.1}{\spb1.2\spb2.3\spb3.4\spb4.1} \,.
\label{R3mmmmb}
\eea
For higher multiplicities, and excluding the amplitudes with three
negative-helicity gravitons, the analytic expressions get unwieldy.
Nevertheless, double copies can be obtained easily once a
(possibly numerically determined) set of numerators for the $F^3$
amplitudes is available.  Using the results from the previous section,
we have computed the following gravitational amplitudes:
\bea
  M_5^{R+}(1^-,2^-,3^-,4^-,5^+) \,, \nnl
  M_5^{R+}(1^-,2^-,3^-,4^-,5^-) \,, \nnl
  M_6^{R+}(1^-,2^-,3^-,4^-,5^+,6^+) \,, \nnl 
  M_6^{R+}(1^-,2^-,3^-,4^-,5^-,6^+) \,, \nnl
  M_6^{R+}(1^-,2^-,3^-,4^-,5^-,6^-) \,.
  \label{availableR3amps}
\eea

\subsection{Consistency checks}

With the set of amplitudes~(\ref{R3MHV}) and (\ref{availableR3amps})
at one's disposal, one can perform various consistency checks.
An immediate check is the total symmetry
under exchange of any pair of equal-helicity gravitons.  This symmetry
is not manifest in the double-copy construction, but we verified it for
all amplitudes, either analytically or numerically.

Another set of relations to be satisfied by the gravitational amplitudes can
be derived from collinear and soft factorization.
For example, suppose the external
momenta $k_4$ and $k_5$ are collinear, obeying $k_4\approx zk_P$ and
$k_5\approx (1-z)k_P$ with $k_P=k_4+k_5$ and $z\in[0,1]$.  Then the following
relation has to hold:
\bea
  M_6^{R+}(1^-,2^-,3^-,4^-,5^+,6^+) &\stackrel{k_4||k_5}{\longrightarrow}&
  \text{Split}_-(4^-,5^+) \, M_5^{R+}(1^-,2^-,3^-,P^+,6^+)
\nnl &&\null
+ \text{Split}_+(4^-,5^+)\, M_5^{R+}(1^-,2^-,3^-,P^-,6^+) \,. \quad
  \label{CollinearRelGrav}
\eea
The gravitational splitting amplitudes applicable to the
above situation are~\cite{BernGaugeGravity}
\bea
 \text{Split}_-(4^-,5^+) &=& - \frac{(1-z)^3}{z} \frac{\spa4.5}{\spb4.5} \,,\nnl
 \text{Split}_+(4^-,5^+) &=& - \frac{z^3}{1-z} \frac{\spb4.5}{\spa4.5} \,.
  \label{GravSplittingExample}
\eea
In other factorization channels one also needs
\bea
 \text{Split}_+(4^-,5^-) &=& - \frac{1}{z(1-z)} \frac{\spa4.5}{\spb4.5} \,,\nnl
 \text{Split}_-(4^-,5^-) &=& 0 \,,
\label{GravSplittingOther}
\eea
plus the complex conjugate splitting amplitudes.
(See ref.~\cite{BernGaugeGravity}, section 5, for subtleties
related to testing the collinear limits of gravity amplitudes.)
The real collinear limits of the `$R^+$' amplitudes have exactly the same
form as in Einstein gravity, because the three-point amplitudes
in this theory, \eqn{3ptKLT}, vanish in the real collinear limit.  Hence
the collinear singularities come just from the Einstein triple-graviton
vertex.

Similarly, the soft limits will involve the standard soft-graviton factor
for Einstein gravity.  An example of a soft-limit relation is
\be
 M_5^{R+}(1^-,2^-,3^-,4^-,5^+)
\ \stackrel{k_4\rightarrow 0}{\longrightarrow}\
 \text{Soft}(4^-) \, M_4^{R+}(1^-,2^-,3^-,5^+)\,,
\ee
where the soft factor in the above equation is given by 
\be
 \text{Soft}(4^-) =
 \frac{\spa1.4\spb1.3\spb2.4\spb5.1 + \spa2.4\spb1.4\spb2.3\spb5.2}
      {\spb1.4\spb2.4\spb3.4\spb5.4} \,,
\ee
and a general form can be found in ref.~\cite{BernGaugeGravity}.

We tested all possible collinear and soft limits for the amplitudes
in \eqn{availableR3amps}. In addition, we tested that the 
collinear and soft limits of our formula~(\ref{R3MHV}) for
amplitudes with three negative-helicity gravitons
are consistent up to the ten-point level.  In short, the set of amplitudes
obtained by double-copying the $F^3$ amplitudes satisfies all expected
collinear and soft limits for multi-graviton amplitudes.

While the agreement between the result of the double-copy (DC)
construction and the application of the field-theory KLT relations
${\rm KLT}_{{\rm FT},m}$ was already noted in
\eqns{3ptKLT2}{4ptR3MHVKLT} for three- and four-point cases,
we have compared the results of the two methods numerically for all
other available amplitudes through six points.
For all amplitudes that we tested, we find
\be
{\rm DC}[A^{F^+},A^{F^+}]\ =\ {\rm KLT}_{\rm FT}[A^{F^+}, A^{F^+}].
\label{DCequalsKLTforFplus}
\ee
Thus for these cases the double-copy construction is equivalent to
applying the field-theory limit of the KLT relations to the $F^3$
amplitudes.  By parity we also have, for the conjugate helicity
configuration,
\be
{\rm DC}[A^{F^-},A^{F^-}]\ =\ {\rm KLT}_{\rm FT}[A^{F^-}, A^{F^-}].
\label{DCequalsKLTforFminus}
\ee
In one of the cases that we tested, the two towers in 
\fig{fig:F3amps} overlap, and an $F^3$ amplitude can receive
nonzero contributions from both $A^{F^+}$ and $A^{F^-}$.
This helicity configuration, in the six-graviton case, is
$({-}{-}{-}{+}{+}{+})$.  In this case it is also of interest
to test for the off-diagonal double-copy or KLT construction.
We find that in this case both constructions vanish:
\be
{\rm DC}[A^{F^+},A^{F^-}]\ =\ {\rm KLT}_{\rm FT}[A^{F^+}, A^{F^-}]\ =\ 0\,,
\label{DCequalsKLTequalszero6plusminus}
\ee
for $({-}{-}{-}{+}{+}{+})$.  It would be interesting to investigate
whether this is true more generally.


\subsection{KLT relations and effective actions for the bosonic string}

Next we address the question as to which gravitational action produces
the `$R^+$' amplitudes from the previous subsection.
The fact that the double-copy construction yields results equivalent to
those obtained by employing the field-theory KLT relations clearly
suggests that the amplitudes may originate from corrections to the
Einstein-Hilbert action that appear in the low-energy effective action
of the closed bosonic string.

The operator $F^3$ appears at $\CO(\a')$ in the low-energy effective
action of the open bosonic string.
Thus, the resulting gravitational amplitudes from double-copying an
amplitude with one insertion of $F^3$ should appear at $\CO(\a'^2)$ in a
gravitational effective action. The low-energy effective action for the closed
bosonic string reads~\cite{MT},
\bea
S&=&-\frac{2}{\k^2}\int \diff^4 x
\sqrt{g}\Big[R-2(\pd_\m\p)^2-\frac{1}{12}H^2\nnl
&&\null\qquad\qquad\qquad\quad + \a' \, \frac{1}{4} \, e^{-2\p} \, G_2 \nnl
&&\null\qquad\qquad\qquad\quad + \a'^2 \, e^{-4\p}
\Big(\frac{1}{48}I_1+\frac{1}{24}G_3\Big)+\CO(\a'^3)\Big] \,,
\label{leeaclosed}
\eea
where the variable $\phi$ denotes the dilaton and $H=dB$ is the
outer derivative of the totally antisymmetric tensor $B_{\m\n}$. In addition,
$G_2=R_{\m\n\r\s}R^{\m\n\r\s}-4R_{\m\n}R^{\m\n}+R^2$ is the usual topological
Gauss-Bonnet contribution with two powers of the Riemann tensor.  The two
terms appearing at $\CO(\a'^2)$ are defined by
\be
I_1=R^{\m\n}_{\;\;\;\;\a\b}R^{\a\b}_{\;\;\;\;\s\r}R^{\s\r}_{\;\;\;\;\m\n}
\qquad\text{and}\qquad 
G_3=I_1-2R^{\m\n}_{\;\;\;\;\a\b}R^{\n\s}_{\;\;\;\;\b\gamma}R^{\s\m}_{\;\;\;\;\gamma\a}\,.
\ee
Both of the $\CO(\a'^2)$ terms in the above equation, $I_1$ and $G_3$,
could in principle result in corrections at $\CO(\a'^2)$.
However, expanding the terms around flat
Minkowski space, one can show~\cite{MT} that only the first term, $I_1$, 
contributes to $\CO(\a'^2)$ terms in the four-graviton amplitudes in which
we are interested (although the $G_3$ term does contribute to mixed
dilaton-graviton four-point amplitudes).

Below, we will refer to the particular combination of operators at
$\CO(\a'^2)$ in the \eqn{leeaclosed} simply as $R^3$.  As we will show
below, the vertices originating from the operators appearing at
$\CO(\a'^2)$ are insufficient to yield the set of amplitudes obtained
from double-copying the amplitudes linear in $F^3$. One needs, in addition,
to consider contributions from two insertions of the vertices appearing at
$\CO(\a')$ in \eqn{leeaclosed}, which are generated by the operator
$e^{-2\phi} G_2$.

An immediate obstacle for a direct comparison between the double-copy results
and amplitudes resulting from the action~(\ref{leeaclosed}) 
is the availability of amplitudes resulting from an insertion of the
operator $R^3$.  While the
implementation of CSW-like rules for amplitudes in the $F^3$ sector above was
straightforward (see~\sect{sec:F3}), the situation in gravity is more
subtle: an attempt to generalize the CSW-formalism to gravity resulted in
complicated rules requiring BCFW-like shifts~\cite{EmilCSW} which renders
calculations cumbersome. As an alternative, one can use all-line
BCFW-shifts~\cite{RisagerCSWProof,EFKProofMHVVertexExpansion}.
However, this method is limited to particular helicity amplitudes that vanish
as the complex shifted momentum becomes large; it cannot be applied to all
helicity configurations.

Let us start with the available amplitudes.  At the three-point level,
the operator $R^3$ generates nonzero amplitudes for the three-graviton
helicity configurations $({-}{-}{-})$ and $({+}{+}{+})$, while
the configurations $({\mp}{\mp}{\pm})$ receive no contribution from it.
For convenience, we will normalize the operator $R^3$ so that its
three-graviton amplitudes are given by $M_3^{R+}$ in \eqn{3ptKLT}.
At the level of four gravitons, Cohen, Elvang and Kiermaier~\cite{CEK}
calculated the all-minus amplitude for pure gravity with one insertion of a
vertex originating in $R^3$.  An anti-holomorphic all-line shift leads to
only one contributing BCFW diagram, with two cubic vertices and a graviton
exchange, and then a sum over all possible permutations of the external
states, $\CP(1234)$, yielding\footnote{%
Compared to the result in ref.~\cite{CEK}, our conventions 
lead to an additional prefactor of $i$ in \eqn{4ptmmmmR3CEK}.}
\be
M_4^{R^3,{\hbox{\small{all-line shift}}}}(1^-,2^-,3^-,4^-)\ =\ 
\frac{i}{4}\sum_{\CP(1234)}
 \frac{\<12\>^5\<34\>^2}{[12]}\frac{[1\xi]^2[2\xi]^2}{[3\xi]^2[4\xi]^2} \,.
\label{4ptmmmmR3CEK}
\ee
We can simplify this expression to
\bea
M_4^{R^3,{\hbox{\small{all-line shift}}}}(1^-,2^-,3^-,4^-) &=&
10 \, i \, stu 
\frac{\spa1.2\spa2.3\spa3.4\spa4.1}{\spb1.2\spb2.3\spb3.4\spb4.1}
\label{4ptmmmmR3CEKa}\\
&=& \frac{5}{2} M_4^{R+}(1^-,2^-,3^-,4^-) \,,
\label{4ptmmmmR3CEKb}
\eea
where $M_4^{R+}$ is the four-graviton amplitude constructed from either
the double-copy formula or the KLT relations, and given in \eqn{R3mmmmb}.

The proportionality of the $R^3$ amplitude to the result from applying
field-theory KLT relations to the all-minus four-point $F^3$ amplitudes
was already noted in ref.~\cite{CEK}.  However, the two expressions are not
equal; there is an prefactor of $\sfrac{5}{2}$ in front of $M_4^{R+}$
which needs to be explained.   The explanation is that the double-copy
construction (or equivalently the field-theory KLT relations)
includes diagrams with axion and dilaton exchange, not just the pure-graviton
contributions in the computation~(\ref{4ptmmmmR3CEK}).\footnote{We thank
Tim Cohen, Henriette Elvang and Michael Kiermaier for this
explanation.}

The real scalar dilaton $\phi$ and pseudoscalar axion $a$ are given in 
terms of a complex scalar field $\varphi$ and its complex conjugate
$\bar\varphi$ as (following the conventions of ref.~\cite{EKStringyKLT})
\be
\phi=\frac{1}{2}(\varphi+\bar{\varphi})\qquad\text{and}\qquad
a=\frac{i}{2}(\varphi-\bar{\varphi}),
\label{dilaxdef}
\ee
respectively. Amplitudes involving $\varphi$ and $\bar\varphi$
can be obtained via the double-copy construction or KLT relations by
combining amplitude pairs whose corresponding gluons have opposite helicity,
\be
\varphi = g^- \otimes g^+
\qquad\text{and}\qquad \bar\varphi = g^+ \otimes g^-\,.
\ee
At $\CO(\a'^2)$, there are contributions to the amplitude with four
negative-helicity gravitons from exchanges of a dilaton or axion between
two three-point vertices, each generated by an insertion of
the operator $e^{-2\p}G_2$ in \eqn{leeaclosed}.

The necessary three-point vertices, with two gravitons and either $\varphi$ or
$\bar\varphi$, are easy to work out by expanding the operator $e^{-2\p}G_2$.
They are simplest to normalize (relative to the $R^3$ amplitude) by using the 
KLT relations.  The results are
\bea
M_3(1^-,2^-,3^\varphi)\ =\ M_3(1^-,2^-,3^{\bar\varphi})
&=&{\rm KLT}_{{\rm FT},3}[A_3^{F+}(1^-,2^-,3^\mp),A_3(1^-,2^-,3^\pm)]\nnl
&=&-i\, \<12\>^4\,.
\label{0mm}
\eea
Angular-momentum conservation implies a vanishing result for the case of
opposite helicity gravitons:
\be
M_3(1^-,2^+,3^\varphi)\ =\ M_3(1^-,2^+,3^{\bar\varphi}) = 0 \,.
\label{0mp}
\ee
The results~(\ref{0mm}) and (\ref{0mp}) are also valid when the
scalar particle is off-shell.  We remark that these three-point vertices
all vanish in the limit of real collinear kinematics.  Therefore the
collinear limits we studied in the previous subsection can receive
no contributions from terms generated by two insertions of $e^{-2\p}G_2$.

Next we compute the scalar exchange contribution to the all-minus
four-graviton amplitude.  We connect two vertices of the form~(\ref{0mm})
with a scalar propagator, include a factor of two to account for both
$\varphi$ and $\bar\varphi$ exchange, and sum over all three exchange
channels, to obtain
\be
M_4^{(\phi R^2)^2}(1^-,2^-,3^-,4^-)\ =\ -2i\,\Bigg(
\frac{\<12\>^4\<34\>^4}{s_{12}}
+\frac{\<13\>^4\<24\>^4}{s_{13}}+\frac{\<23\>^4\<41\>^4}{s_{23}} 
\Bigg)\,.
\label{4ptmmmmaxdil}
\ee
(In the dilaton/axion basis~(\ref{dilaxdef}), all the contributions
come from the dilaton; the diagrams from axion exchange cancel.)
The amplitude in \eqn{4ptmmmmaxdil} turns out to be precisely
$-\sfrac{3}{2}$ times the KLT result~(\ref{R3mmmmb}).

In summary, the KLT expression~(\ref{R3mmmmb}) can be decomposed
as,
\bea
M_4^{R+}(1^-,2^-,3^-,4^-) &=& 
M_4^{R^3,{\hbox{\small{all-line shift}}}}(1^-,2^-,3^-,4^-) +
M_4^{(\p R^2)^2}(1^-,2^-,3^-,4^-)~~~~~~
\label{mmmmdecomp1}\\
&=& \left( \frac{5}{2} \, - \, \frac{3}{2} \right)
M_4^{R+}(1^-,2^-,3^-,4^-)\,.
\label{mmmmdecomp2}
\eea
This result shows that applying the double-copy construction to a pair of
amplitudes, each with one insertion of a vertex originating in $F^+$,
yields graviton amplitudes with gravitons and dilaton/axions on internal
lines.  Although this was the simplest possible example, it is not
difficult to verify this behavior for other amplitudes, 
at least when the pure $R^3$ contribution is available.
For example, we have compared the double-copy result
for the five-point all-minus amplitude with an all-line-shift
calculation of the $R^3$ contribution, and we find the expected result:
\bea
M_5^{R+}(1^-,2^-,3^-,4^-,5^-) &=&
  M_5^{R^3,{\hbox{\small{all-line shift}}}}(1^-,2^-,3^-,4^-,5^-)
+ M_5^{(\p R^2)^2}(1^-,2^-,3^-,4^-,5^-)\nonumber\\
&=& \left( \frac{5}{2} \, - \, \frac{3}{2} \right)
M_5^{R+}(1^-,2^-,3^-,4^-,5^-)\,.
\label{mmmmmDecomp}
\eea
The second contribution on the right-hand side of the first equation
contains a four-point subdiagram with dilaton exchange
(the axion exchange contribution again vanishes).  For general
amplitudes, the $R^3$ and dilaton/axion contributions presumably
need not be proportional, although they are in the two all-minus
cases we computed explicitly, \eqns{mmmmdecomp2}{mmmmmDecomp}.

For the four-graviton helicity configuration $({-}{-}{-}{+})$,
the vanishing of the opposite-helicity three-point amplitude
$M_3^{\CO(\a')}(1^-,2^+,3^\varphi)$ in \eqn{0mp} implies that the 
KLT-constructed amplitude $M_4^{R+}(1^-,2^-,3^-,4^+)$ given in
\eqn{4ptR3MHVb} should have no contributions from dilaton/axion exchange.
That is, it should come purely from $R^3$:
\be
 M_4^{R+}(1^-,2^-,3^-,4^+) = M_4^{R^3}(1^-,2^-,3^-,4^+)\,.
\label{mmmpEquality}
\ee
We can test these results against a Feynman-diagram computation by
van Nieuwenhuizen and Wu~\cite{NieuWu} of the pure-graviton $R^3$
amplitudes for the helicity configurations
$({-}{-}{-}{-})$ and $({-}{-}{-}{+})$.\footnote{The result in
ref.~\cite{NieuWu}
is written in a convention where momenta are assumed to be ingoing for ingoing
particles and outgoing for outgoing particles.}
In order to avoid discussing phases and operator normalization conventions,
we take the magnitude of the ratio of amplitudes they computed,
\be
\left| \frac{M_4^{\rm vNW}(1^-,2^-,3^-,4^-)}{M_4^{\rm vNW}(1^-,2^-,3^-,4^+)} \right|
= \left| \frac{\sfrac{15}{4}i stu}{\sfrac{3}{8}istu}\right|
\label{vNWratio}
= 10 \,.
\ee
Our corresponding ratio, from eqs.~(\ref{4ptR3MHVb}), (\ref{R3mmmmb})
and (\ref{4ptmmmmR3CEKb}), is,
\be
\frac{5}{2} \left| \frac{M_4^{R+}(1^-,2^-,3^-,4^-)}
                        {M_4^{R+}(1^-,2^-,3^-,4^+)} \right|
= \frac{5}{2} \times \frac{4stu}{stu} = 10,
\label{ourratio}
\ee
in perfect agreement with \eqn{vNWratio}.

The final four-graviton helicity configuration to consider at $\CO(\a'^2)$
is $({-}{-}{+}{+})$.  There is no contribution to this amplitude
from one insertion of the operator $R^3$~\cite{NieuWu}.
Also, in the effective action of the closed superstring there is no
contribution at this order (the first pure-graviton correction is at
$\CO(\a'^3)$ from the $R^4$ operator).  Finally, there is no
contribution from the KLT-square of two $F^3$ amplitudes,
because the component amplitudes vanish,
\be
M_4^{R+}(1^-,2^-,3^+,4^+) \, \equiv \,
{\rm KLT}_{{\rm FT},4}[ A_4^{F^+}(1^-,2^-,3^+,4^+),A_4^{F^+}(1^-,2^-,3^+,4^+) ]
\, = \, 0 \,.  
\label{mmppNoKLT}
\ee
On the other hand, there is a nonzero contribution in the
closed bosonic string, and this contribution can be accounted for
by two insertions of $e^{-2\p}G_2$.  In this case, we just need to
connect the vertex~(\ref{0mm}) for $M_3(1^-,2^-,3^\varphi)$
with its relabeled complex-conjugate,
\be
M_3^{\p R^2}(1^+,2^+,3^{\varphi}) = -i \, {\spb1.2}^4 \,,
\label{0pp}
\ee
via a scalar propagator in the $s$ channel.  Again there are
two contributing diagrams (from $\varphi$ and $\bar\varphi$ exchange),
but this time only one of the three channels contributes.
Thus one finds for the four-point gravity amplitude,
\be
M_4^{(\phi R^2)^2}(1^-,2^-,3^+,4^+)
= - 2\,i\,\frac{\<12\>^4[34]^4}{s_{12}} \,.
\label{4ptmmppdilexch}
\ee
Again the result all comes from dilaton exchange; the axion-exchange
diagrams cancel. We used the KLT relations~\cite{KLT} to check that the
$\CO(\a'^2)$ correction to the closed bosonic string tree-level amplitude
indeed agrees with \eqn{4ptmmppdilexch}. 

\Eqn{4ptmmppdilexch} is different from the previous examples in the
following respect:  In all previous cases, the amplitudes `$M_m^{R+}$'
generated by squaring $F^3$ amplitudes could be generated consistently from
the $\CO(\a'^2)$ terms in the closed bosonic string effective action.
But in this case it is not true, because $M_4^{R+}(1^-,2^-,3^+,4^+)$
vanishes according to \eqn{mmppNoKLT}, while $M_4^{(\phi R^2)^2}$ does not.
This discrepancy can be accounted for when we realize that at $\CO(\a'^2)$
in the string-theory KLT relations there are also contributions from
Taylor expanding the sine functions to third order, and utilizing the leading
YM terms in both open-string amplitudes.  These contributions account
for \eqn{4ptmmppdilexch} in the context of the KLT relations.

All of the above discussion was based on squaring amplitudes with one 
insertion of a vertex originating in $F^3$, for both the left- and
right-moving gauge theory.  However, the double-copy construction,
as well as the KLT relations, in principle allow one to combine amplitudes
originating in different theories and with different symmetries,
provided that a color-kinematic dual representation is available for
at least one of
the two copies. For example, one- and two-loop amplitudes in $\CN>4$
supergravities have been constructed in this way~\cite{BoucherDixon}, 
and even more remarkably, the (vanishing) ultraviolet divergence in 
$\CN=4$ supergravity at three loops~\cite{Bern2012cd}.

Because dual representations are available for all gluon tree amplitudes in
pure Yang-Mills theory (and we have found the same to be true for all 
gluon amplitudes with one insertion of an $F^3$ vertex), we can combine YM 
with $F^3$ amplitudes in order to obtain pure-graviton amplitudes which
would be at $\CO(\a')$.  However, we expect to get a vanishing result,
because of an absence of potential operators, after field redefinitions
and taking into account that the Gauss-Bonnet term is a total derivative
in four dimensions.

For three and four external gravitons, the vanishing is trivial, because
for those multiplicities there are no common nonzero helicity structures
between pure YM and $F^3$ amplitudes, as has been shown in~\sect{sec:F3}.
However, starting from the five-point level,
helicity configurations with nonzero amplitudes of both kind exist.
For example, the usual pure gauge-theory \MHV{} helicity
configuration $({-}{-}{+}{+}{+})$ coincides with the \MHVbar{} configuration
from the $F^3$ sector.  Thus, one can straightforwardly square those
amplitudes into pure graviton ones.  We have checked that the graviton
amplitudes resulting from this mixed-KLT construction vanish for all
helicity configurations at five and six points:
\be
{\rm KLT}_{{\rm FT},m}[ A_m, \tA_m^{F+} ]\ =\ 0 \,, \qquad m=5,6.
\label{FTKLTorthog}
\ee
This vanishing result is just as expected from the form of the
$\CO(\a')$ term in \eqn{leeaclosed} \,,
\be
e^{-2\phi}G_2 = \Bigl[ 1 - 2\phi + \CO(\phi^2) \Bigr] G_2 \,,
\ee
because $G_2$ is a total derivative in four dimensions. 
Because it is purely topological, it cannot contribute to an amplitude. 
However, if one allows for external dilatons, this is not the case: a
term like $-2\phi G_2$ cannot be written as a total derivative and thus will
contribute on shell, as we see already in the three-point
amplitudes~(\ref{0mm}) and~(\ref{0pp}). 

Thus the two sets of amplitudes, YM and $F^3$, are an example of a pairing,
in which both sets satisfy color-kinematics duality, but the `off-diagonal'
double-copy construction, or KLT construction~(\ref{FTKLTorthog}),
does not yield nontrivial pure-graviton amplitudes.
On the other hand, it does generate mixed scalar-graviton amplitudes,
such as the three-point amplitude $M_3(1^-,2^-,3^\varphi)$ in \eqn{0mm}.

In summary, from the point of view of computing general $R^3$
amplitudes in gravity, the double-copy construction based on two sets
of $F^3$ amplitudes does provide one ingredient.  However, it would have
to be combined with a separate computation of the effects of double
insertions of $\phi R^2$ with dilaton exchange, as well as the effects of
the YM-squared terms in the string-theory KLT relations, after expanding
the sine functions to third order.

\section{Beyond $F^3$}\label{sec:Beyond}

In this section, we briefly discuss the prospects for finding
color-kinematic dual representations or BCJ amplitude relations
for operators at dimension eight, built from four powers of the
gluon field strength tensor $F_{\m\n}$.  Once one combines more than three
powers of the field strength, the index contraction pattern is no longer
unique.  At the next order, $F^4$ for short, there are four different
contractions available
(see {\it e.g.}~refs.~\cite{TseytlinVEffAction,BjB2003a}):
\bea
   &&\Tr(F_\m^{~\n} F_\n^{~\r} F_\r^{~\s} F_\s^{~\m}),
 \,\,\Tr(F_\m^{~\n}  F_\r^{~\s} F_\n^{~\r} F_\s^{~\m})\nnl
   && \Tr(F_{\m\n}  F_{\r\s} F^{\m\n} F^{\r\s}),
 \,\,\Tr(F_{\m\n} F^{\m\n} F_{\r\s} F^{\r\s})\,.
   \label{F4options}
\eea
(As at dimension six in pure YM, terms with covariant derivatives are
removable by using field redefinitions.)
In pure gauge theory, any of these insertions can occur separately with
an independent coefficient. However, if the theory is supersymmetric, the
possible coefficients are constrained.  The particular unique linear
combination appearing at $\CO(\a'^2)$ in the low-energy effective action of 
the open superstring (ss), corresponding to $\CN=4$ supersymmetry,
is~\cite{GreenSchwarz,SchwarzSuperstringTheory,TseytlinVEffAction}:
\bea
  F^4_{\text{ss}} &\equiv& 
  \, \Tr\Bigl[ F_\m^{~\n} F_\n^{~\r} F_\r^{~\s} F_\s^{~\m}
  + 2 F_\m^{~\n}  F_\r^{~\s} F_\n^{~\r} F_\s^{~\m} \nnl
  &&\hskip0.7cm\null
   - \textstyle{\frac{1}{4}} F_{\m\n}  F_{\r\s} F^{\m\n} F^{\r\s}
   - \textstyle{\frac{1}{2}} F_{\m\n}  F^{\m\n} F_{\r\s} F^{\r\s} \Bigr] \,.
  \label{F4StringTheory}
\eea

Amplitudes originating from insertions of operators appearing in
the string effective action can be derived from the infinite-tension limit
$\a'\rightarrow 0$ of the corresponding string amplitude. After factoring out
the field-theory MHV amplitude, the expansion for the four-gluon amplitude
reads~\cite{StiebergerTaylor}\footnote{%
In ref.~\cite{StiebergerTaylor} another metric signature is used,
which accounts for the sign difference with respect to the expression there.}:
\be
  A_{\text{ss}}(1^-,2^-,3^+,4^+)\ =\
   i \, \frac{{\spa1.2}^4}{\spa1.2\spa2.3\spa3.4\spa4.1}
 \, \bigl[ 1-\a'^2\zeta(2)st+\a'^3\zeta(3)stu + \CO(\a'^4) \bigr] \,,
  \label{4ptStringExpansion}
\ee
while expressions for higher multiplicity are available from the same
reference. A short calculation reveals that amplitudes at
$\CO(\a'^2)$ in the superstring expansion do not allow for a
color-kinematic dual representation. In order to find the reason for
this behavior, let us consider the string-theory monodromy relations
once again~\cite{MonodromyStieberger,MonodromyNBI1,MonodromyNBI2}.
For simplicity we will write the relations at four points, however,
higher multiplicities work analogously. The four-point form of
\eqn{Monodromy1} relating color-ordered subamplitudes of the string
tree-level amplitude reads:
\begin{equation}
  A^{\rm string}(1,2,3,4)+e^{i\pi\a' s}A^{\rm string}(2,1,3,4)
+e^{i\pi\a'(s+u)}A^{\rm string}(2,3,1,4) = 0.
\end{equation}
In order to obtain relations between different parts of the string amplitude
appearing at different orders in $\a'$, one needs to expand the string
amplitude $A^{\rm string}$ via
\begin{equation}
  A^{\rm string}=A^{\YM}+\a'A^{\a'}+\a'^2A^{\a'^2}+\a'^3A^{\a'^3}+\cdots\,.
\end{equation}
Also expanding the exponential, one finds that the monodromy
relations at the first few orders in $\a'$ read\footnote{%
For completeness we have also included the relations already appearing
in \eqns{Monodromy2a}{Monodromy2}.}: 
\begin{eqnarray}
  \CO(\a'^0):\quad 0&=&A^{\YM}(1,2,3,4)+A^{\YM}(2,1,3,4)+A^{\YM}(2,3,1,4),
  \label{ME1}\\
  \CO(\a'^1):\quad 0&=&s \, A^{\YM}(2,1,3,4)- t \, A^{\YM}(2,3,1,4),
 \label{ME2}\\
  \CO(\a'^1):\quad 0&=&A^{\a'}(1,2,3,4)+A^{\a'}(2,1,3,4)+A^{\a'}(2,3,1,4),
 \label{ME3}\\
 \CO(\a'^2):\quad  0&=&s \, A^{\a'}(2,3,1,4)-t \, A^{\a'}(2,1,3,4),
\label{ME4}\\
 \CO(\a'^2):\quad  0&=& - \frac{\pi^2}{2} s^2 \, A^{\YM}(2,3,1,4)
                       - \frac{\pi^2}{2} t^2 \, A^{\YM}(2,1,3,4)
\label{ME5}\nnl
 &&\null + A^{\a'^2}(1,2,3,4)+A^{\a'^2}(2,1,3,4)+A^{\a'^2}(2,3,1,4), \\
 \CO(\a'^3):\quad  0&=& - \frac{\pi^2}{6} s^3 \, A^{\YM}(2,1,3,4)
                        + \frac{\pi^2}{6} t^3 \, A^{\YM}(2,3,1,4)
\label{ME6}\nnl
 &&\null + s \, A^{\a'^2}(2,1,3,4) - t \, A^{\a'^2}(2,3,1,4), \\
 \CO(\a'^3):\quad  0&=& - \frac{\pi^2}{2} s^2 \, A^{\a'}(2,1,3,4)
                       - \frac{\pi^2}{2} t^2 \, A^{\a'}(2,3,1,4)
\label{ME7}\nnl
 &&\null + A^{\a'^3}(1,2,3,4)+A^{\a'^3}(2,1,3,4)+A^{\a'^3}(2,3,1,4).
  \label{MonodromyExpansion}
\end{eqnarray}
These relations are valid for supersymmetric as well as bosonic string
theory.  In the former case, the expressions can be simplified by 
noticing that $A^{\a'}$ vanishes. 

The first two lines in the above expansions, \eqns{ME1}{ME2}, yield 
the photon decoupling and BCJ relations for the YM amplitudes,
while \eqns{ME3}{ME4} manifest the same relations for the amplitudes
$A^{\a'}$, which are equivalent to the $F^3$ amplitudes. \Eqn{ME4}
is one example of the BCJ amplitude representations obtained
for all $F^3$ amplitudes in~\sect{sec:F3}.

The second part of \eqn{ME5} has the same form as the photon-decoupling 
equation in the lines above, but for $A^{\a'^2}$.  However, this relation
is spoiled by the two other terms, given by Mandelstam
variables multiplying the YM amplitudes.  Thus, one should not expect the
amplitudes at $\CO(\a'^2)$ in the bosonic string or the superstring
to satisfy the usual photon-decoupling identity or BCJ relations
({\it i.e.}~\eqn{ME6} also has extra terms).

Indeed, considering the collection of
\eqns{CubicAntisymmetry}{4ptJac}, together with the amplitude
equations, {\it e.g.},
\be
A^{\a'^2}(1,2,3,4)\ =\ \frac{n_s}{s}+\frac{n_t}{t} \,,
  \label{Asp2tring}
\ee
and the corresponding ones for $A^{\a'^2}(1,2,4,3)$ and $A^{\a'^2}(3,1,4,2)$, the
system does not have nontrivial solutions for any helicity structure, neither
in bosonic nor in superstring theory.

Is there a way to explain this behavior in terms of the color structure
accompanying the corresponding higher-derivative operators at the
appropriate orders in  $\a'$?   For the simple example of the four-point 
YM tree amplitude ($\CO(\a'^0)$) in~\sect{sec:Background},
the symmetries of the kinematic part allowed rewriting the trace-based
amplitude decomposition in terms of the totally antisymmetric structure
constants $f^{abc}$ only.  (Of course this must be possible, since
the YM Feynman rules are expressed in this way.)

For the insertion of the unique dimension-six operator $F^3$ this property
was already shown at the level of the action (see \eqn{totalantisymF3}):
One can decompose a trace of three matrices into symmetric and antisymmetric
parts,
\be
\Tr(T^a \, T^b \, T^c)=d^{abc}+\frac{1}{2} \tf^{abc}\,.
  \label{Tr3Expansion}
\ee
Here $d^{abc}$ is the case $n=3$ of the more general totally symmetric trace
structure, summed over all $n!$ permutations $\CP(a_1,\ldots,a_n)$,
\be
  d^{a_1\dots a_n}
 = \frac{1}{n!}\sum_{\s\in\CP(a_1,\ldots,a_n)}\Tr(T^{a_{\s(1)}} \ldots T^{a_{\s(n)}}) \,.
  \label{symmetrictrace}
\ee
However, for $F^3$ the antisymmetry of the field strength $F_{\m\n}$
removes the symmetric part, leaving only the totally antisymmetric
part $\tf^{abc}$. 

In contrast, for the open superstring at $\CO(\a'^2)$ the operator
$F^4_{\text{ss}}$ in \eqn{F4StringTheory} projects onto the totally symmetric
part of the space of four-color-matrix traces~\cite{Bilal,TseytlinBI}:
\begin{equation}
  \Tr(T^aT^bT^cT^d) \, (F^4_\text{ss})_{abcd}
\ \propto\ d^{abcd} \, (F^4_\text{ss})_{abcd} \,,
\end{equation}
where the general color-trace can be decomposed as,
\be
  {\rm Tr}(T^a T^b T^c T^d)\ =\ d^{abcd}
  + \frac{1}{2}(\tf^{bce}d^{ead}-\tf^{ade}d^{ebc})
  - \frac{1}{6}(\tf^{ade}\tf^{ebc}-\tf^{abe}\tf^{ecd}) \,.
\label{Tr4Expansion}
\ee
It is not possible to write $d^{abcd}$ purely in terms of antisymmetric
structure constants $\tf^{abc}$.  Because the Kleiss-Kuijf and, in particular,
the photon-decoupling identity are proven by writing amplitudes in
terms of structure constants, it is no surprise that the amplitudes
$A^{\a'^2}$ generated by $F^4_\text{ss}$ do not even satisfy photon-decoupling,
\eqn{ME5}.  Therefore this combination of $F^4$ operators is not
color-compatible with a color-kinematics duality based on cubic graphs.

In order to search for combinations of $F^4$ operators that are 
color-compatible, we first recognize that \eqns{ME5}{ME6} are valid for the
open bosonic string (bs) and the open superstring (ss) at the same time, and
that the YM amplitudes are the same in both theories.
Therefore if we take the difference (diff) of the two sets of amplitudes,
\begin{equation}
 A^{\a'^2}_{\rm diff} = A^{\a'^2}_{\rm bs} - A^{\a'^2}_{\rm ss} \,,
\end{equation}
then the differences of the corresponding \eqns{ME5}{ME6} will remove the
common YM terms that were spoiling the photon-decoupling and BCJ relations
at $\CO(\a'^2)$:
\begin{eqnarray}
 0 &=&  A^{\a'^2}_{\rm diff}(1,2,3,4) + A^{\a'^2}_{\rm diff}(2,1,3,4)
      + A^{\a'^2}_{\rm diff}(2,3,1,4)\label{aprime2photon}\\
 0&=& s \, A^{\a'^2}_{\rm diff}(2,1,3,4) - t \, A^{\a'^2}_{\rm diff}(2,3,1,4) \,.
\label{aprime2bcj}
\end{eqnarray}
Thus, one should be able to express the difference between the bosonic and
superstring effective actions at $\CO(\a'^2)$ in terms of operators
that are color-compatible (built only out of $\tf^{abc}$s).

The two obvious $F^4$-type operators with color structure $\tf^{abe}\tf^{cde}$
are
\begin{equation}
  O_A=\Tr([F_\m^{~\n}, F_\n^{~\r}]\,[F_\r^{~\s}, F_\s^{~\m}]) 
\quad\text{and}\quad O_B=\Tr([F_{\m\n}, F_{\r\s}]\,[ F^{\m\n}, F^{\r\s}])\,.
\end{equation}
These operators are linear combinations of the operators given in
\eqn{F4options}.  Their contributions to four-point amplitudes can be
determined easily.
Amplitudes generated by both operators necessarily satisfy the
photon-decoupling identity.  On the other hand, neither operator,
nor any linear combination of $O_A$ and $O_B$,
generates amplitudes that satisfy the BCJ amplitude relations.
Thus a color-kinematic dual representation is impossible.
How can this be, given that \eqn{aprime2bcj} looks like a BCJ relation?

The explanation is similar to that for the closed-string discrepancy between
$M_4^{R+}(1^-,2^-,3^-,4^-)$ and $M_4^{R^3}(1^-,2^-,3^-,4^-)$ discussed
in~\sect{sec:R3}:  One has to take into account two insertions
of a lower-dimension operator.
In this case, it is the operator $F^3$ at order $\a'$, and gluons
are exchanged between the two insertions.  Performing the appropriate
Feynman-diagram calculation leads to a third contribution at $\CO(\a'^2)$,
denoted by $A^{(F^3)^2}$.  Not surprisingly, the contribution $A^{(F^3)^2}$
alone satisfies photon-decoupling, but not the BCJ relations. 
However, a specific linear combination of $A^{O_A}$, $A^{O_B}$ and $A^{(F^3)^2}$
correctly reproduces the amplitude difference $A^{\a'^2}_{\rm diff}$
between the bosonic string and the superstring:
\begin{equation}
   A^{\a'^2}_{\rm bs} - A^{\a'^2}_{\rm ss}
 = A^{\a'^2}_{\rm diff}
 = \frac{1}{2}A^{(F^3)^2} + \frac{1}{4} A^{O_B} \,.
   \label{bsap2}
\end{equation}
Expanding the four-point amplitude for the bosonic string
(Veneziano amplitude) to $\CO(\a'^2)$ and subtracting the
superstring result~(\ref{4ptStringExpansion}) gives,
\bea
  A^{\a'^2}_{\rm diff}(1^-,2^-,3^+,4^+)
  &=& - ut \, A^\YM(1^-,2^-,3^+,4^+)\,, \\
  A^{\a'^2}_{\rm diff}(1^-,2^+,3^-,4^+)
  &=& - st \, A^\YM(1^-,2^+,3^-,4^+)\,.
\eea
The corresponding contributions from $(F^3)^2$ and $O_B$ are
given by,
\bea
\frac{1}{2}A^{(F^3)^2}(1^-,2^-,3^+,4^+) &=&  
\frac{t^2-ut}{2} \, A^\YM(1^-,2^-,3^+,4^+)\,,\\
\frac{1}{4} A^{O_B}(1^-,2^-,3^+,4^+) &=& 
\frac{st}{2} \, A^\YM(1^-,2^-,3^+,4^+) \,, \\
\frac{1}{2}A^{(F^3)^2}(1^-,2^+,3^-,4^+) &=&
0\,, \\\
\frac{1}{4} A^{O_B}(1^-,2^+,3^-,4^+) &=& 
- st \, A^\YM(1^-,2^+,3^-,4^+) \,. 
\eea

Using the fact that $stA^\YM$ is totally symmetric, and
factoring this quantity out, it is easy to see that $A^{\a'^2}_{\rm diff}$
satisfies photon decoupling,
\bea
&&A^{\a'^2}_{\rm diff}(1^-,2^-,3^+,4^+) + A^{\a'^2}_{\rm diff}(2^-,1^-,3^+,4^+)
      + A^{\a'^2}_{\rm diff}(2^-,3^+,1^-,4^+)\nonumber\\
&=& stA^\YM(1^-,2^-,3^+,4^+) \left[ \frac{u}{s} + \frac{t}{s} + 1 \right]
 = 0\,,
\eea
and the BCJ relation,
\be
s \, A^{\a'^2}_{\rm diff}(2^-,1^-,3^+,4^+) - t \, A^{\a'^2}_{\rm diff}(2^-,3^+,1^-,4^+)
= stA^\YM(1^-,2^-,3^+,4^+) \left[ s \, \frac{t}{s} - t  \right] = 0\,.
\ee
Similarly, $A^{O_B}$ obeys photon decoupling,
\bea
&&A^{O_B}(1^-,2^-,3^+,4^+) + A^{O_B}(2^-,1^-,3^+,4^+) + A^{O_B}(2^-,3^+,1^-,4^+)
\nonumber\\
&=& stA^\YM(1^-,2^-,3^+,4^+) \left[ 2 + 2 - 4 \right]
 = 0\,,
\eea
but the BCJ relation does not hold,
\be
s \, A^{O_B}(2^-,1^-,3^+,4^+) - t \, A^{O_B}(2^-,3^+,1^-,4^+)
= stA^\YM(1^-,2^-,3^+,4^+) [ 2\,s - 4\, t ] \neq 0\,.
\ee
The above example shows that the correct color structure alone is not a
sufficient condition for the existence of a color-kinematic dual
representation.  In addition, one needs the right kinematic behavior,
which in this case could only happen if one combines two insertions of the
dimension-six operator $F^3$ and one insertion of the dimension-eight
operator $O_B$ with the right relative coefficient.

We also tested higher-order in $\a'$ contributions to the superstring
amplitude.  Except for $\CO(\a'^3)$, we have not found color-kinematic
dual representations to exist for even the four-point amplitude.
At $\CO(\a'^3)$, the four-gluon amplitude does obey the photon-decoupling
and BCJ relations.  Therefore we have examined the five- and six-gluon
amplitudes as well, using the open-string amplitudes from
refs.~\cite{StiebergerTaylor,Stieberger2007am};
we find that they obey both sets of equations.  The
effective action at this order is a particular linear combination of
operators of the type $D^2F^4$ and $F^5$, where $D$ denotes a
gauge-covariant derivative~\cite{Bilal}.

\begin{table}[!ht]
\begin{tabular}{|>{$}c<{$}|>{$}c<{$}|}
  \hline
  a^{(0)}&-\frac{1}{2s}\Big(\frac{1}{u}\tf^{ace}\tf^{bde}
           +\frac{1}{t}\tf^{ade}\tf^{bce}\Big)\tabularnewline
\hline a^{(1)}&0\tabularnewline
\hline a^{(2)}&-6\zeta(2) d^{abcd}\tabularnewline
\hline a^{(3)}&-\frac{\zeta(3)}{2}(u\tf^{abe}\tf^{cde}+s\tf^{ace}\tf^{bde})
\tabularnewline
\hline a^{(4)}&-\frac{3}{2}\zeta(2)^2(s^2+t^2+u^2)d^{abcd}\tabularnewline
&\quad-\frac{\zeta(4)}{8}[
 s(u-t)\tf^{abe}\tf^{cde}+u(s-t)\tf^{ace}\tf^{bde}+t(s-u)\tf^{ade}\tf^{bce}]
\tabularnewline\hline
\end{tabular}
\caption{Corrections to the tree-level four-point gauge-theory
  amplitude from the low-energy expansion of open superstring
  theory~\cite{Bilal}. }
\label{colortab}
\end{table}

These results nicely match the known color structure of the four-point
amplitude~\cite{Bilal}.  The four-point superstring amplitude can be
represented as
\bea
  \CA(1,2,3,4)|_{\a'\rightarrow 0} = K(1,2,3,4) \sum_{n=0}^\infty \a'^n a^{(n)} \,,
  \label{BilalExpansion}
\eea
where the prefactor $K(1,2,3,4)$ describes the complete state
dependence and $a^{(n)}$ is a totally symmetric contribution
containing the color dependence at $\CO(\a'^n)$. Considering the
factors in table \ref{colortab} from ref.~\cite{Bilal} (which
contains results through $\CO(\a'^6)$),
one can see that (with the exception of $\CO(\a'^3)$) they are not
purely built from $\tf^{abc}$s, but contain totally symmetric $d^{abcd}$
contributions.  Thus it is not surprising that one cannot find tree-level
dual representations for stringy $\a'$ corrections beyond $\CO(\a'^3)$.
Finally, we note that the $\CO(\a'^3)$ case, for which we do have evidence
of color-kinematics
duality through at least six external states, does not suffer from
multiple insertions of lower-dimension operators, because there is no
$\CO(\a')$ correction to the superstring amplitude.  The first case
of multiple insertions is delayed until $\CO(\a'^4)$.

Another area in which one might search for color-kinematics duality
is in `form factors', that is, the matrix elements of operators $O(x)$
with plane-wave states in gauge theory.  In momentum space, generic momentum
is injected at the operator.  Perhaps the simplest case to consider
is $O = \Tr(F^{\m\n}F_{\m\n})$ or $\tilde{O} = \Tr(F^{\m\n}\tilde{F}_{\m\n})$.
The injected momentum can be thought of as carried by
a scalar or pseudoscalar particle.  For $O$ (or $\tilde{O}$), the
coupling to gluons is the same as for the Standard Model Higgs boson 
(or its pseudoscalar partner $A$ in a two-Higgs doublet model)
in the heavy top-quark limit, through
the coupling $H\Tr(F^{\m\n}F_{\m\n})$ (or $A\Tr(F^{\m\n}\tilde{F}_{\m\n})$). 
Matrix elements for a Higgs boson coupling to multiple gluons have been
studied often, for example in ref.~\cite{DixonGloverKhoze}.

Here we simply wish to ask whether these matrix elements might exhibit
color-kinematics duality. At first sight, one might be optimistic,
because the class of MHV amplitudes,
\be
A_m(\phi,1^+,2^+,\ldots,p^-,\ldots,q^-,\ldots,m^+)\,,
\label{PhiAmps}
\ee
where $\phi=\sfrac{1}{2}(H+iA)$, has precisely the same
formula~\cite{DixonGloverKhoze}
as the pure YM Parke-Taylor amplitudes~(\ref{PTformula}).
However, when one checks the BCJ amplitude relation for four gluons,
using this formula, one finds that it is spoiled by the momentum
injected by the $\phi$ particle.  The photon-decoupling relation continues
to hold, however, for all values of $m>2$.  The latter holds because the
color-dressed $\phi$-amplitudes ${\cal A}_m(\phi,\ldots)$ for $m>2$ can
be expressed solely in terms of structure constants~$\tf^{abc}$.

However, the $m=2$ amplitude is exceptional:  It is 
proportional to $\delta^{a_1a_2}$,
\be
{\cal A}_2(\phi,1^-,2^-)\ \propto\ \delta^{a_1a_2} {\spa1.2}^2 \,.
\ee
Hence the color-ordered amplitude $A_2(\phi,1^-,2^-)$ does not even
obey photon-decoupling.  Heuristically, we can attribute the failure
of the BCJ amplitude relations for $m>2$ to the mismatch between
color and kinematics in a cubic-graph representation:
There should be a $\phi gg$ kinematical vertex to account for the momentum
injected by $\phi$, but there is no corresponding $\tf^{abc}$
color factor.


\section{Conclusions and outlook}\label{sec:Conclusions}

In this article we studied whether color-kinematics duality in gauge theory
can be extended beyond the renormalizable level, deforming the pure YM 
theory by higher-dimension operators composed from the gluon field strength.
Using string-theory monodromy relations, we found that the amplitudes
produced by one insertion of the operator $F^3$ obey BCJ amplitude relations.
We verified these relations for a variety of multiplicities and helicity
configurations, using CSW rules to construct the amplitudes (in perfect
numerical agreement with a Feynman-diagram calculation using {\sc MadGraph}).
We also determined (often numerically) the existence of color-kinematic
dual representations for all these cases.

The double copy of those representations yields amplitudes which originate in
corrections to Einstein gravity present in the effective action of the closed
bosonic string up to order $\CO(\a'^2)$. Besides diagrams involving an
$R^3$ operator and gravitons on internal lines only, one also has to
consider dilaton-exchange diagrams involving two insertions of a $\phi R^2$
operator.  The effects of YM-squared terms in the string-theory KLT relations,
after expanding the sine functions of the momentum invariants 
to third order, have to be considered as well in some cases.

Our results show --- up to the order tested --- that the double-copy
construction can be equivalent to the KLT relations at higher orders in the 
inverse string tension $\a'$.  In addition to squaring the modified
amplitudes and checking the gravity amplitudes for Bose symmetry,
collinearity and soft limits, we implemented all-line BCFW shifts for
suitable $R^3$ gravity amplitudes as an additional test. 

For several other additions of operators to the Yang-Mills action,
tree-level color-kinematic dual
representations do not exist.  In some cases we could
attribute this property to the color structure of the operator.  If
the operator is not composed entirely from totally antisymmetric
structure constants $f^{abc}$ (as in examples drawn from the
low-energy effective action of open string theory), then a standard
dual representation will not exist.  In fact, the
photon-decoupling identities will not be obeyed.  Whether a
generalized color-kinematics duality could hold in some of these
cases, involving four- or higher-index structure
constants~\cite{Bargheer2012gv}, is a question that is certainly
deserving of future investigation.  Another avenue for further work
could be to investigate higher-dimension operators in gauge theory
that include matter fields as well as gauge fields.

We also found that even some operators that are color-compatible with
the usual color-kinematics duality do not actually lead to amplitudes
obeying it, such as the dimension-eight $F^4$-type operators $O_A$ and $O_B$
studied in \sect{sec:Beyond}.  Thus, color structure alone is not sufficient 
to ensure color-kinematics duality.  However, one of these operators, $O_B$,
can be combined with two insertions of $F^3$ in order to restore
color-kinematics duality.  This combination corresponds to
the difference of terms in the bosonic and superstring effective actions.
Also, the $\CO(\a'^3)$ term in the superstring effective action
appears to generate amplitudes consistent with color-kinematics duality,
at least through six external gluons.
The above results hint that color-kinematics duality may be closely
related to string theory also beyond the leading order in $\a'$; however,
more detailed work would be needed to fully explore these connections.
The recent motivic organization of superstring
amplitudes~\cite{Schlotterer2012ny}
may provide very useful clues in this regard.


\section*{Acknowledgments}

We thank Zvi Bern, Rutger Boels, John Joseph Carrasco, Tim Cohen,
Henriette Elvang, Michael Kiermaier and Radu Roiban for stimulating
discussions and correspondence.  J.B. is indebted to Claude Duhr for a
lightning introduction to the use of {\sc MadGraph} and for providing
the {\sc MadGraph} implementation for $F^3$ vertices via {\sc FeynRules}.
This work was supported in part by the US Department of
Energy under contract DE--AC02--76SF00515.  J.B. gratefully
acknowledges support from the Alexander-von-Humboldt foundation within
the Feodor-Lynen program.

\end{document}